\documentclass[aps,prl,superscriptaddress,twocolumn]{revtex4}

\usepackage{graphicx,chemarr,color,hyperref,soul}
\usepackage[version=4]{mhchem}
\usepackage{cancel}
\usepackage{amssymb,amsmath,textgreek}
\begin{document}

\title{Precision of Protein Thermometry}

\author{Michael Vennettilli}
\affiliation{Department of Physics and Astronomy, University of Pittsburgh, Pittsburgh, PA, 15260, USA}
\affiliation{Department of Physics and Astronomy, Purdue University, West Lafayette, Indiana 47907, USA}

\author{Soutick Saha}
\affiliation{Department of Physics and Astronomy, Purdue University, West Lafayette, Indiana 47907, USA}

\author{Ushasi Roy}
\affiliation{Department of Physics and Astronomy, Purdue University, West Lafayette, Indiana 47907, USA}

\author{Andrew Mugler}
\email{andrew.mugler@pitt.edu}
\affiliation{Department of Physics and Astronomy, University of Pittsburgh, Pittsburgh, PA, 15260, USA}
\affiliation{Department of Physics and Astronomy, Purdue University, West Lafayette, Indiana 47907, USA}

\begin{abstract}
Temperature sensing is a ubiquitous cell behavior, but the fundamental limits to the precision of temperature sensing are poorly understood. Unlike in chemical concentration sensing, the precision of temperature sensing is not limited by extrinsic fluctuations in the temperature field itself. Instead, we find that precision is limited by the intrinsic copy number, turnover, and binding kinetics of temperature-sensitive proteins. Developing a model based on the canonical TlpA protein, we find that a cell can estimate temperature to within 2\%.
We compare this prediction with {\it in vivo} data on temperature sensing in bacteria.
\end{abstract}

\maketitle

Cells routinely make decisions based on the temperature of their surroundings. For example, most cells undergo systemic changes in response to a heat or cold shock \cite{mccarty_dnak_1991}. Some cells initiate a phenotypic response such as virulence when the temperature crosses a particular threshold \cite{falconi_thermoregulation_1998}. Some cells thermotax, or move toward a preferred temperature range \cite{maeda_effect_1976}. These behaviors are possible because molecular conformations, chemical reaction rates, and various mechanical properties of cells can change dramatically as a function of temperature, and cells have developed many different ways to detect such changes \cite{schumann_thermosensors_2007, klinkert_microbial_2009, mandin_feeling_2020}. Molecules that participate in the response to temperature changes are called molecular thermometers or thermosensors, and this class includes DNA and various RNA and protein molecules.

Despite detailed knowledge of the molecular mechanisms of temperature sensing in cells, the basic question of what sets the precision of temperature sensing remains largely unexplored. Is the precision limited extrinsically by temperature fluctuations in the surrounding fluid, or intrinsically by properties of the cell's molecular components? Similar questions have been heavily investigated for other types of cell sensing, beginning with Berg and Purcell's analysis of chemical concentration sensing \cite{berg_physics_1977}, and extending to sensing of concentration gradients \cite{endres_accuracy_2008}, concentration ramps \cite{mora_limits_2010}, multiple ligands \cite{mora_physical_2015}, material stiffness \cite{beroz_physical_2017}, and fluid flow \cite{fancher_precision_2020}, among others. In most of these cases, extrinsic fluctuations have been found to limit sensory precision, suggesting that cells have evolved sensors that are as precise as physically possible. However, the precision of temperature sensing, and the associated question of extrinsic versus intrinsic limits, has been understudied by comparison.

Early work by Dusenbery shed important light on this problem \cite{dusenbery_limits_1988}. Using the two-point correlation function for temperature fluctuations in a homogeneous fluid, Dusenbery estimated that extrinsic fluctuations are several orders of magnitude smaller than cells' actual sensitivity thresholds. This finding suggests that cells' temperature sensors are not as precise as physically possible. However, it leaves an important question unanswered: if extrinsic fluctuations do not set the limit on the precision of cellular temperature sensing, then what does?

Here we revisit this problem from a perspective that combines the physics of temperature fluctuations with the molecular mechanisms of thermoreception. Following Dusenbery's lead, we start by using the two-point correlation function to investigate a thermal analog of Berg and Purcell's ``perfect instrument" for concentration sensing \cite{berg_physics_1977}. This investigation confirms that extrinsic temperature fluctuations are far too small to be limiting in a biological context. We therefore investigate the intrinsic fluctuations imposed by cells' molecular machinery for temperature sensing. We are guided by a prototypical and well studied protein thermometer, namely the TlpA protein in the bacterium \textit{Salmonella typhimurium} \cite{koski_new_1992, hurme_dna_1996, hurme_proteinaceous_1997}. Developing a stochastic model based on the experimentally characterized details of TlpA, we find that intrinsic fluctuations are much larger than extrinsic fluctuations and can in fact be biologically limiting. Specifically, we find that intrinsic fluctuations impose a sensing error of roughly 2\%,
and we discuss how this limit compares with the observed temperature sensing threshold in bacteria.

\begin{figure}
	\includegraphics[width=1\columnwidth]{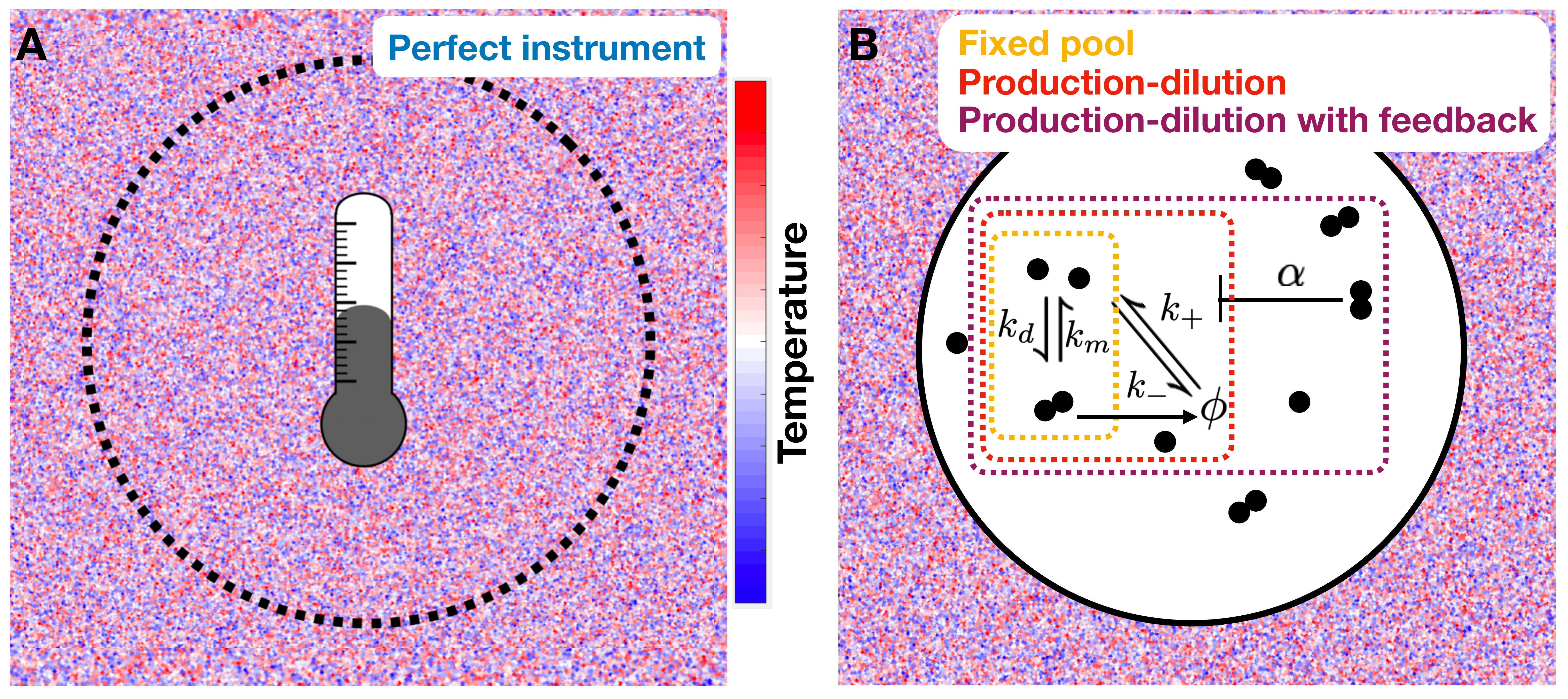}
	\caption{Temperature sensing (A) via an analog of Berg and Purcell's \cite{berg_physics_1977} perfect instrument for concentration sensing, and (B) via a protein thermometer. Based on the TlpA protein, monomers reversibly dimerize, monomers are expressed, monomer and dimers are diluted by cell division, and dimers inhibit monomer expression.}
	\label{fig:models}
\end{figure}

In their perfect instrument for concentration sensing, Berg and Purcell considered a completely permeable sphere of radius $a$ that could count the number of molecules within its volume at each instant, perform a time average, and use this information to estimate the surrounding concentration \cite{berg_physics_1977}.
In the case of temperature sensing, the analogous instrument is a permeable sphere of radius $a$ that records the temperature $T(\vec{x},t)$ at each point within its volume at each instant $t\in[0,\tau]$, performs a volume and time average, and then uses the result as the temperature estimate $\hat{T}$ (Fig.\ \ref{fig:models}A).
We assume the medium to be homogeneous and in thermal equilibrium, with average temperature $\overline{T}$. The key ingredient is the two-point correlation function for the temperature fluctuations obtained in the regime of linear irreversible thermodynamics \cite{fox_gaussian_1978}
\begin{equation}\label{eq:tempCovar}
\begin{aligned}
&\langle(T(\vec{x},t) -\overline{T})(T(\vec{x}\,',t')-\overline{T})\rangle  \\
&= \frac{k_B \overline{T}^2}{\rho c_s} \left(\frac{\rho c_s}{4\pi K|t-t'|}\right)^{3/2}
\exp \left[ -\frac{\rho c_s ||\vec{x}-\vec{x}\,'||^2}{4K|t-t'|} \right],
\end{aligned}
\end{equation}
where $k_B$ is Boltzmann's constant and the material properties $\rho$, $c_s$, and $K$ are the mass density, specific heat, and thermal conductivity of the medium respectively.
The variance in the estimator is computed by integrating the two-point correlation function in Eq.\ \ref{eq:tempCovar} in both space and time. The result has the following short- and long-time limits \cite{supp},
\begin{equation}\label{eq:extrinsic}
	\frac{\sigma(\hat{T})}{\overline{T}} = \sqrt{\frac{k_B}{C}} \times
	\begin{cases}
	\qquad 1 & \tau\rightarrow 0 \\
	\sqrt{4\tau_D/(5\tau)} & \tau \gg \tau_D,
	\end{cases}
\end{equation}
where we have introduced the heat capacity of the medium contained within the instrument $C = 4\pi a^3 \rho c_s/3$ and the timescale for temperature fluctuations to diffuse across the instrument $\tau_D = \rho c_s a^2/K$ \cite{fox_gaussian_1978}.
Equation \ref{eq:extrinsic} has an intuitive interpretation: the variance falls off with the heat capacity of the instrument (in units of $k_B$) because if the heat capacity is large, a large fluctuation in thermal energy corresponds to a small fluctuation in temperature. The variance is further decreased in the long-time limit by the number $\tau/\tau_D$ of independent measurements the instrument can make, where independence is defined by the diffusion time.

\begin{figure}
	\includegraphics[width=1\columnwidth]{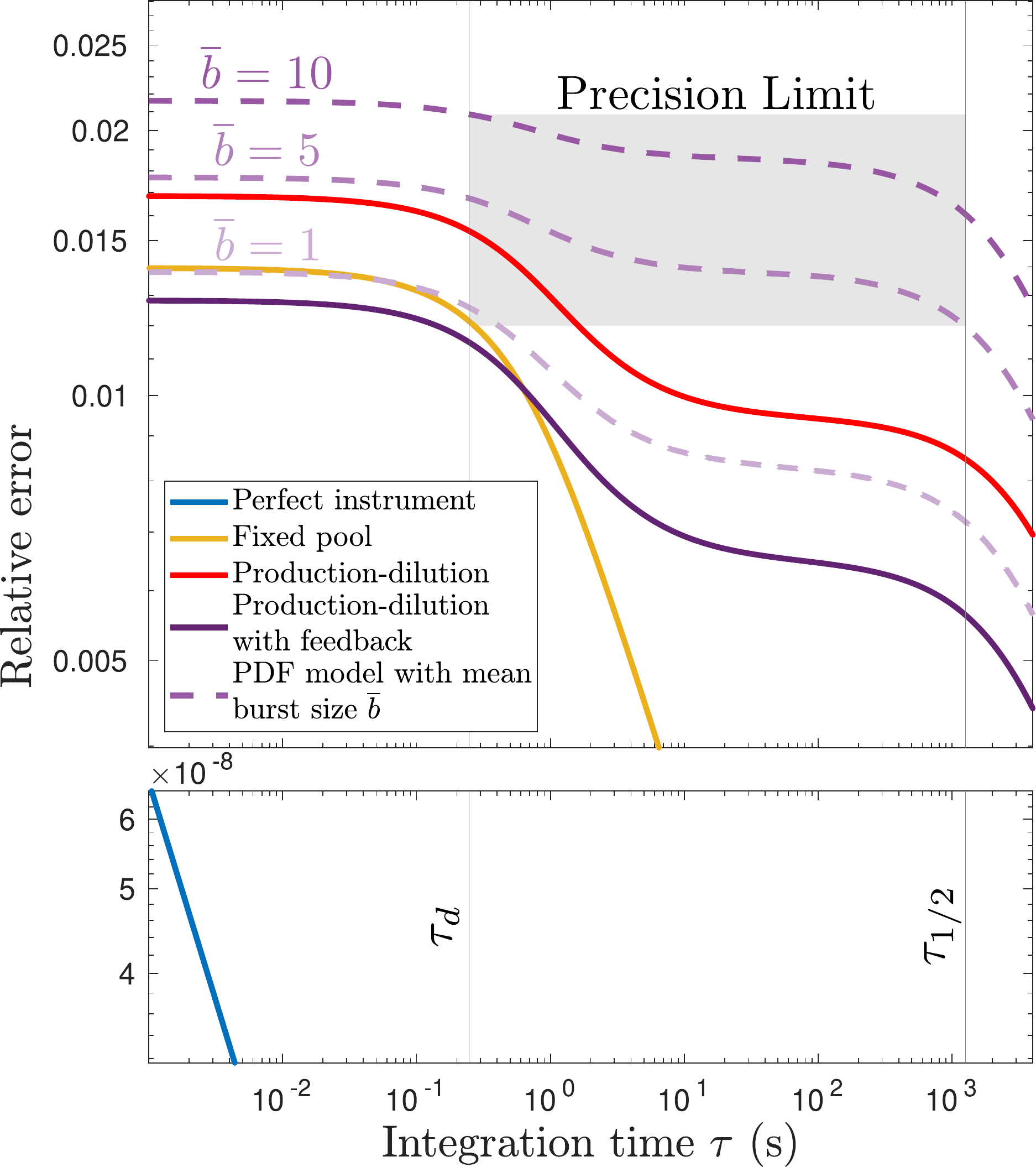}
	\caption{Relative temperature estimation error $\sigma(\hat{T})/\Delta T$ as a function of monomer-number integration time $\tau$. We predict that the error is bounded from below by 2\% (gray box). Parameters are estimated from data as described in the text.
	}
	\label{fig:story}
\end{figure}

For water at room temperature, $\rho \approx 1$ g/cm$^3$, $c_s \approx 4$ J/(g$\cdot$K), and $K \approx 0.6$ J/(s$\cdot$m$\cdot$K). For a cell radius of $a \approx 1$ $\mu$m, the error in an instantaneous measurement according to Eq.\ \ref{eq:extrinsic} is $\sigma(\hat{T})/\overline{T} \approx 10^{-6}$. The diffusion time is $\tau_D \approx 6$ $\mu$s, after which the error drops further due to time averaging (Fig.\ \ref{fig:story}, blue). Clearly the extrinsic fluctuations in the medium itself are not limiting, as it is unlikely that a cell needs to estimate temperature to less than one part in a million. This finding agrees with the conclusions of Dusenbery, whose approach was more heuristic \cite{dusenbery_limits_1988}.

Of course, cells are not perfect thermometers. They detect temperature indirectly through molecular or mechanical properties \cite{schumann_thermosensors_2007, klinkert_microbial_2009, mandin_feeling_2020}. Therefore, to investigate the intrinsic limits imposed by the detection mechanism itself, we must develop a model that accounts for the information actually available to the cell. Here we focus on the molecular mechanism of protein thermometry, in which proteins' conformational states are temperature dependent. Protein thermometers are ubiquitous: for example, temperature-dependent oligomerization, unfolding or misfolding, and methylation of proteins drive, in various combinations, the heat shock response \cite{mccarty_dnak_1991}, high-temperature response \cite{hurme_proteinaceous_1997, servant2000rhea}, and thermotaxis response \cite{maeda_effect_1976,jiang_mechanism_2009,paulick_thermo_2017} in bacteria. In these cases, a temperature-induced conformational change is generally followed by negative feedback \cite{note_feedback}.

For concreteness, we consider the protein TlpA in \textit{S.\ typhimurium}, which includes these common features but is otherwise relatively simple and experimentally well characterized.
A step-increase in temperature results in a sustained increase in TlpA level, suggesting that TlpA responds to absolute temperature, not temperature change \cite{note_absolute}.
TlpA forms homodimers, with the dimer favored at low temperatures and the monomer favored at high temperatures \cite{koski_new_1992, hurme_proteinaceous_1997}. The dimer binds to the promoter region of the \textit{tlpA} gene and inhibits its expression \cite{hurme_dna_1996}, resulting in negative feedback.
TlpA is a canonical protein thermometer, and its mechanism has been used to engineer other thermal switches \cite{piraner_tunable_2017,piraner_modular_2019}.

Suppose that two TlpA molecules associate with rate $k_d$ and dissociate with rate $k_m$ \cite{koski_new_1992} (Fig.\ \ref{fig:models}B, yellow). Subject to these reactions alone, the total number of TlpA units $n = m + 2d$ is conserved, where $m$ and $d$ are the numbers of monomers and dimers respectively. Therefore we refer to this as the ``fixed pool" (FP) model.
The mean fraction $f = \overline{m}/n$ of TlpA units in the monomeric state has been measured as a function of temperature at physiological concentrations using circular dichroism spectroscopy \cite{hurme_proteinaceous_1997,heyn_circular_1975}. We find that the data are well described by a sigmoid $f(T) = \{1+ \exp[-4(T-T_M)/\Delta T]\}^{-1}$ with half-maximal temperature $T_M =39\ ^{\circ}$C and width $\Delta T= 6.3\ ^{\circ}$C \cite{supp} [the factor of 4 ensures that $f'(T_M) = 1/\Delta T$]. We assume that the cell infers the temperature from the mean monomer number, which it estimates from the time average $\hat{m}_{\tau} = \tau^{-1} \int_0^{\tau}m(t)dt$ \cite{berg_physics_1977} (we find similar results if temperature is instead inferred from the dimer number \cite{supp}). In the supplement we also consider maximum likelihood estimation \cite{cover_elements_2012}, which in this case has the least squared error of all possible estimators, and find that it performs similarly to the naive time average considered here \cite{supp}.

To convert the error in monomer number estimation to that in temperature estimation, we use linear error propagation \cite{berg_physics_1977}, $\sigma(\hat{T}) = \sigma(\hat{m}_{\tau})/|d\overline{m}/dT| = \sigma(\hat{m}_{\tau})/(nf')$, where the second step follows from $\overline{m}=nf$.
To find $\sigma(\hat{m}_{\tau})$, we perform the second-order Kramers-Moyal expansion and linearize to obtain the fluctuations \cite{van_kampen_stochastic_2011, gardiner_stochastic_2009, klebaner_introduction_2012}. The result is \cite{supp}
\begin{equation}\label{eq:fixedPool}
\frac{\sigma(\hat{T})}{\Delta T} = \frac{\sigma_{\rm FP}(m)}{nf'\Delta T}\times
\begin{cases}
\qquad 1 & \tau\rightarrow 0 \\
\sqrt{2\tau_d/\tau} & \tau \gg \tau_d,
\end{cases}
\end{equation}
where $\sigma^2_{\rm FP}(m) =  2nf(1-f)/(2-f)$ is the instantaneous variance in the monomer number and $\tau_d = c(k_d\overline{m})^{-1}$ is the autocorrelation time, with $c = (1-f)/[2(2-f)]$ a numerical factor \cite{note_DeltaT}.
Equation \ref{eq:fixedPool} has an intuitive interpretation: the factor $nf(1-f)$ in $\sigma_{\rm FP}^2(m)$ is the variance of the binomial distribution, which arises because the molecules switch between the monomer and dimer states. The additional factor $2/(2-f)$ is an increase in the noise due to the fact that dimerization further discretizes the monomer number beyond that of a pure binomial process, as the monomer number can only change by two \cite{roob_cooperative_2016}. Finally, $(k_d\overline{m})^{-1}$, which sets $\tau_d$, is the timescale for a monomer to form a dimer with any other monomer. As in Eq.\ \ref{eq:extrinsic}, the variance in the long-time limit of Eq.\ \ref{eq:fixedPool} is reduced by the number $\tau/\tau_d$ of independent measurements made.

When $T=T_M$, we have $f=1/2$ and $f' = 1/\Delta T$, and the instantaneous error in Eq.\ \ref{eq:fixedPool} reduces to $\sigma(\hat{T})/\Delta T = 1/\sqrt{3n}$. We see that the error decreases with the square root of the number of TlpA molecules $n$, as expected for counting noise.
From the experimentally estimated number of TlpA dimers per cell \cite{hurme_proteinaceous_1997}, we infer $n\approx 1700$ \cite{note_dimer}, and therefore an instantaneous error of $\sigma(\hat{T})/\Delta T = 1.4\%$ (Fig.\ \ref{fig:story}, yellow). 
To see how sensing improves with time integration, we need to estimate the dimerization rate $k_d$. We are unaware of an experimental estimate for the dimerization rate of TlpA. However, TlpA is a coiled-coil, and the dimerization rate of engineered coiled-coils has been measured at $k_dV = 4\times10^5$ (M$\cdot$s)$^{-1}$ \cite{chao_use_1998}. Given the bacterial volume of $V=1\ \mu$m$^3$ \cite{smit_outer_1975}, this results in an autocorrelation time of $\tau_d = 0.3$ s at $f=1/2$, beyond which the error falls off \cite{code}. The intrinsic noise from molecular detection (Fig.\ \ref{fig:story}, yellow) clearly dominates over the extrinsic noise from temperature fluctuations in the medium (Fig.\ \ref{fig:story}, blue).

The fixed pool model is unrealistic because in cells the protein number is not actually fixed. Instead, proteins are produced via gene expression and lost by active degradation or dilution from cell division. As we are not aware of evidence that TlpA is actively degraded, we consider dilution here.
Specifically, we introduce a production rate $k^+$ for the monomer and a dilution rate $k^-$ for both the monomer and dimer.
We call this the ``production-dilution'' model
(Fig.\ \ref{fig:models}B, red).
Experiments \cite{hurme_proteinaceous_1997, https://doi.org/10.1046/j.1365-2672.1998.00410.x} suggest that neither $k^+$ nor $k^-$ is strongly temperature dependent \cite{supp}, and therefore we assume that the dominant temperature dependence is via $f$.
Because cell division is much slower than monomer binding \cite{milo_cell_2015}, we consider the limit $k^- \ll k_d \overline{m}$.

Using the same stochastic techniques as above, we find \cite{supp} that the mean and variance of the monomer number become $\overline{m} = fk^+/k^-$ and
\begin{equation}
\label{eq:fullVar}
	\sigma^2(m) = \sigma^2_{\text{FP}}(m) + \frac{f^2 \sigma^2(n)}{(2-f)^2},
\end{equation}
where $\sigma^2(n)= (7-3 f) k^+/(4 k^-)$ is the variance of the (now fluctuating) pool size $n=m+2d$, and $\sigma^2_{\text{FP}}(m)$ as given beneath Eq.\ \ref{eq:fixedPool} is here written in terms of the mean pool size $\overline{n} = \overline{m}/f$. The second term in Eq.\ \ref{eq:fullVar} is always positive, showing that pool fluctuations due to protein turnover increase the noise, as expected. 
Indeed, using $f=1/2$ and $k^+/k^-$ inferred from the experimental dimer number
\cite{note_dimer},
 we see that the instantaneous error (Fig.\ \ref{fig:story}, red) is increased from that of the FP model (Fig.\ \ref{fig:story}, yellow).
The full $\tau$-dependent expression for $\sigma(\hat{T})/\Delta T$ is calculated
\cite{code} using $k^- = \ln(2)/\tau_{1/2} \approx 2$ hr$^{-1}$ from cell division \cite{lowrie_division_1979}, and we see that the relative error has two clear bends at the dimerization and dilution timescales $\tau_d$ and $\tau_{1/2}$ respectively
(Fig.\ \ref{fig:story}, red).

Thus far we have not yet accounted for the fact that TlpA exhibits negative feedback: the TlpA dimer binds to the promoter region of the \textit{tlpA} gene and inhibits its expression \cite{hurme_dna_1996}.
To incorporate this autorepression, we replace the monomer production rate $k^+$ with the function $k^+/(1+\alpha d)$. We call this the ``production-dilution with feedback'' (PDF) model (Fig.\ \ref{fig:models}B, purple). The parameter $\alpha$ describes the autorepression strength, and its inverse sets where half-maximal expression occurs. Experiments \cite{hurme_proteinaceous_1997} suggest that $\alpha$ is not strongly temperature dependent \cite{supp}, and therefore we continue to assume that the dominant temperature dependence is via $f$.
With autorepression, we find \cite{supp} that the mean monomer number becomes
\begin{equation}
\label{eq:fullMean}
	\overline{m} = \frac{f}{\alpha(1-f)}\left[\sqrt{1+ \frac{2\alpha k^+(1-f)}{k^-}}-1\right], \\
\end{equation}
and the variance obeys Eq.\ \ref{eq:fullVar} with $\sigma^2(n)$ acquiring an $\alpha$ dependence (see \cite{supp}).
We have checked \cite{supp} that Eqs.\ \ref{eq:fullVar} and \ref{eq:fullMean} agree with stochastic simulations \cite{gillespie_exact_1977}.
Both Eq.\ \ref{eq:fullVar} and Eq.\ \ref{eq:fullMean} decrease monotonically with $\alpha$, showing that autorepression reduces both the monomer number variance and its mean. The latter effect dominates, such that relative fluctuations $\sigma(m)/\overline{m}$ increase with autorepression strength \cite{note_alpha}.

The increase in relative fluctuations with autorepression is offset by an increase in temperature sensitivity. To see this, we recognize that the instantaneous relative error can be written $\sigma(\hat{T})/\Delta T = [\sigma(m)/\overline{m}]/[|d\overline{m}/dT|(\Delta T/\overline{m})]$, again by error propagation. The first term in brackets is the relative fluctuations while the second term is the sensitivity: the derivative $d\overline{m}/dT$ scaled by the characteristic quantities $\overline{m}$ and $\Delta T$. Differentiating Eq.\ \ref{eq:fullMean}, the sensitivity evaluates to
\begin{equation} \label{eq:sensitivity}
\frac{d\overline{m}}{dT} \frac{\Delta T}{\overline{m}}
	= \frac{f'\Delta T}{(1-f)}\left[\frac{1}{f}-\frac{1}{2}-\frac{1}{2\sqrt{1+2\alpha k^+(1-f)/k^-}}\right].
\end{equation}
Equation \ref{eq:sensitivity} is an increasing function of $\alpha$, showing that autorepression increases the sensitivity.
This result in consistent with the fact that mutations that target the autorepression result in a weakened dependence of monomer number on temperature \cite{hurme_proteinaceous_1997}.

The tradeoff between increasing relative fluctuations and increasing sensitivity leads to an optimal autorepression strength $\alpha^* = 1.75k^-/k^+$ that minimizes the error in instantaneous temperature sensing $\sigma(\hat{T})/\Delta T$ at $T = T_M$ \cite{supp}. Using this value, $f=1/2$, $k^- = 2$ hr$^{-1}$, and $k^+/k^-$ inferred from the experimental dimer number \cite{note_dimer}, we see that the error (Fig.\ \ref{fig:story}, purple solid) \cite{code} is reduced from the case without feedback (Fig.\ \ref{fig:story}, red).

Finally, we account for a ubiquitous source of additional noise in bacterial gene expression, namely bursts. Bursts of protein production can occur at the transcriptional level, due to binding and unbinding at the promoter region \cite{raj2008nature}, and at the translational level, due to multiple proteins being produced from a single transcript \cite{xie2008single}. In our case the promoter binding timescale is sufficiently fast compared to the protein production timescale that transcriptional bursting can be neglected \cite{supp, erickson_size_2009, 10.1534/genetics.112.143370}, and therefore we focus on translational bursts. Specifically, we perform stochastic simulations \cite{gillespie_exact_1977} of the PDF model in which each production event generates $b$ TlpA proteins instead of one, where $b$ is geometrically distributed with mean $\overline{b}$ \cite{xie2008single}, and we take $k^+ \to k^+/\overline{b}$ to leave the mean monomer number $\overline{m}$ unchanged. We see in Fig.\ \ref{fig:story} that the temperature estimation error increases with mean burst size $\overline{b}$, as expected (purple dashed).

Our results provide a quantitative prediction for the precision with which a cell can estimate temperature using a protein thermometer. A temperature-sensitive behavioral response is likely to occur on a timescale slower than monomer binding $\tau_d$ but faster than cell division $\tau_{1/2}$. Fig.\ \ref{fig:story} shows that the estimation error is relatively insensitive to the integration time in this range. In particular, for a typical bacterial protein burst size of $\overline{b} = 5$$-$$10$ molecules \cite{xie2008single}, we predict that the cell can estimate temperature to within 2\% (Fig.\ \ref{fig:story}, gray box).

How does the predicted bound of $2\%$ precision compare to observed thermosensing thresholds in experimental systems? The transcriptional activity of TlpA has been measured \textit{in vivo} \cite{hurme_proteinaceous_1997} using a Miller assay with a LacZ reporter \cite{miller_experiments_1972, garcia_comparison_2011}. Miller units are proportional to the number of TlpA production events and therefore include time integration while excluding noise downstream of TlpA. Measurements at temperatures $T_1$ and $T_2$ below and above the transition temperature, respectively, provide an estimate of the thermosensing error $\sigma(\hat{T})/\Delta T$, where $\Delta T = T_2-T_1$, and $\sigma(\hat{T})$ is evaluated from the measured uncertainties using linear error propagation (see \cite{supp} for details). Using this procedure, we find $\sigma(\hat{T})/\Delta T = 24\%$. This value is
larger than $2\%$, indicating that this protein thermometer obeys the predicted bound.
In fact, modeling the LacZ reporter explicitly, the predicted bound becomes $20$$-$$30\%$ due to the additional reporter noise \cite{supp}, which is consistent with the experimental observation of 24\%.

The excellent agreement between the predicted bound and the experimental observation may be partly fortuitous. First, the data may include purely experimental sources of error associated with the Miller assay, which would increase the observed error. Second, the Miller assay is a population measurement, which would decrease the observed error: it reports $[\sigma(\hat{T})/\Delta T]/\sqrt{N}$, where $N$ is the number of independently responding units within the population of $N_{\rm cells}$, and the degree to which cells respond in a correlated ($N\to1$) or uncorrelated ($N\to N_{\rm cells}$) manner is unclear. Third, the population likely includes natural cell-to-cell variability \cite{foreman_variability_2020}, which would increase the observed error. These unknowns underscore the need for measurements of temperature sensitivity at the single-cell level. We are not aware of any such measurement for a protein thermometer.

Molecular thermometers drive a variety of cell behaviors, and it is natural to ask how our work could be extended. Many thermosensors, including TlpA, are speculated to cause threshold-like responses, where the cell cares only if the temperature is above a particular threshold, not the value of the temperature itself. For this task, decision theory or optimal stopping \cite{siggia_decisions_2013, berger_statistical_1985, peskir_optimal_2006} may be more appropriate than the time-integrated statistics we investigate here. Furthermore, many thermosensors are used for thermotaxis, the motion of a cell toward an optimal temperature. Here the sensory network is more complicated \cite{jiang_mechanism_2009, paulick_thermo_2017} and the task is also different: the cell cares about the value of both the temperature and its spatial gradient. It would be interesting to integrate our findings into a model of thermotaxis to investigate the physical limits to the precision of that behavior.

Guided by a canonical protein thermometer, we have derived the physical limits to the precision of cellular temperature sensing. Unlike for many other types of cell sensing, the precision of temperature sensing is evidently not limited by the extrinsic noise inherent to the environmental signal itself. Instead, the precision is limited by the biochemical details of the molecular thermometer inside the cell. Specifically, the relative error falls off with the square root of the number of molecules and the number of correlation times, as expected for systems dominated by biochemical noise.
Developing a model based on the experimental features and measured parameters of the TlpA protein, we predict a sensitivity threshold of 2\%, which we find is consistent with the observed thermosensing threshold in bacteria.
Our work advances the understanding of cell sensing and lays the groundwork for further exploration of temperature-sensitive cell behavior.

\acknowledgments
This work was supported by the Simons Foundation (376198) and the National Science Foundation (PHY-1945018).

%\bibliography{ThermoThresh}
%\bibliographystyle{unsrt}

\onecolumngrid
\section{Supplemental Material}

\section{Derivation of Eq.\ 2 of the Main Text}
Fox (Ref.\ [17] of the main text) considered local fluctuations in the temperature $T(x,t)$ of a homogeneous medium near equilibrium with a uniform, time-independent mean
\begin{equation}
\langle T(\vec{x},t)\rangle = \overline{T}.
\end{equation}
This was done in the context of linear, irreversible thermodynamics, and the correlation function for fluctuations in this regime is (Eq.\ 1 of the main text)
\begin{equation}
\langle \Delta T(\vec{x},t) \Delta T(\vec{y},t') \rangle = \frac{k_B \overline{T}^2}{\rho c_s} \left(\frac{\rho c_s}{4\pi K|t-t'|}\right)^{3/2} \exp \left[ -\frac{\rho c_s ||\vec{x}-\vec{y}||^2}{4K|t-t'|} \right],
\end{equation}
where $k_B$ is Boltzmann constant, $\rho$ is the mass density of the medium, $c_s$ is the specific heat, and $K$ is the thermal conductivity. We generalize the perfect instrument of Berg and Purcell to sense temperature as follows. We assume that the detector is a completely permeable sphere of radius $a$ that can record the fluctuating temperature at each point within its volume at each instant. It then performs an average over its volume and some time interval of length $\tau$, yielding the estimate
\begin{equation}
\hat{T} = V^{-1} \tau^{-1} \int_V d^3 \vec{x} \int_0^{\tau} T(\vec{x},t)dt.
\end{equation}
Since it is linear in the temperature, we note that
\begin{equation}
\langle \hat{T} \rangle = \overline{T}.
\end{equation}
The fluctuations are given by the double integral of the correlation function in space and time
\begin{equation}
\sigma^2(\hat{T}) = V^{-2} \tau^{-2} \int_V d^3 \vec{x} d^3 \vec{y} \int_0^{\tau} \langle \Delta T(\vec{x},t) \Delta T(\vec{y},t') \rangle dt dt'.
\end{equation}

\subsection{Short-Time Limit}
In the short-time limit, $\tau\rightarrow 0$, the average is purely spatial
\begin{equation}
\hat{T} \sim V^{-1} \int_V T(\vec{x},t)d^3 \vec{x}
\end{equation}
and the correlation function is a delta function
\begin{equation}
\langle \Delta T(\vec{x},t) \Delta T(\vec{y},t) \rangle = \frac{k_B \overline{T}^2}{\rho c_s} \delta(\vec{x}-\vec{y}).
\end{equation}
With this, the variance in our estimator is
\begin{equation}
\sigma^2(\hat{T})\sim \frac{k_B \overline{T}^2}{\rho c_s V^2} \int_V \delta(\vec{x}-\vec{y}) d^3 \vec{x} d^3 \vec{y} = \frac{3 k_B \overline{T}^2}{4\pi \rho c_s a^3}.
\end{equation}
The noise-to-signal ratio is
\begin{equation}
\frac{\sigma(\hat{T})}{\overline{T}} \sim \sqrt{\frac{3 k_B }{4\pi \rho c_s a^3}},
\end{equation}
as in Eq.\ 2 of the main text (top case).

\subsection{Long-Time Limit}
We will start by performing the time integrals first. We perform a change of variables from $(t,t')$ to $(\Delta,t')$, with $\Delta = t-t'$, and switch the order of integration so that we integrate over $t'$ first. We can do the time integrals analytically, which yields
\begin{equation}
\begin{aligned}
\sigma^2(\hat{T}) = \frac{k_B \overline{T}^2}{V^2} &\int_V \left[ -\frac{e^{-c_s\rho ||\vec{x}-\vec{y}||^2/4K\tau}\sqrt{\rho c_s}}{2(\pi K \tau)^{3/2}} \right. \\
&\left. + \frac{(c_s\rho ||\vec{x}-\vec{y}||^2 + 2K\tau)}{4 \pi K^2 ||\vec{x}-\vec{y}|| \tau^2}\text{erfc}\left(||\vec{x}-\vec{y}||\sqrt{\frac{\rho c_s}{4K\tau}}\right) \right]d^3 \vec{x}d^3 \vec{y}.
\end{aligned}
\end{equation}
In the long-time limit, the term that goes as $\tau^{-1}$ decays the slowest and dominates the expression. We may also set the complementary error function to one, as the argument is small. This simplifies the expression to 
\begin{equation}
\sigma^2(\hat{T}) \sim \frac{k_B \overline{T}^2}{2 \pi K \tau V^2} \int_V \frac{d^3 \vec{x}d^3 \vec{y}}{ ||\vec{x}-\vec{y}||}.
\end{equation}
One can evaluate this integral by expanding it in terms of spherical harmonics or recognizing that it is related to the volume averaged potential inside of a uniformly charged sphere. Either way, the integral comes out to
\begin{equation}
V^{-2}\int_V \frac{d^3 \vec{x}d^3 \vec{y}}{ ||\vec{x}-\vec{y}||} = \frac{6}{5 a}.
\end{equation}
Putting everything together, we find
\begin{equation}
\frac{\sigma(\hat{T})}{\overline{T}} \sim \sqrt{\frac{3 k_B}{5\pi K a \tau}},
\end{equation}
as in Eq.\ 2 of the main text (bottom case).

\section{Fit to the Circular Dichroism Data}

\begin{figure}[b]
	\includegraphics[width=0.5\textwidth]{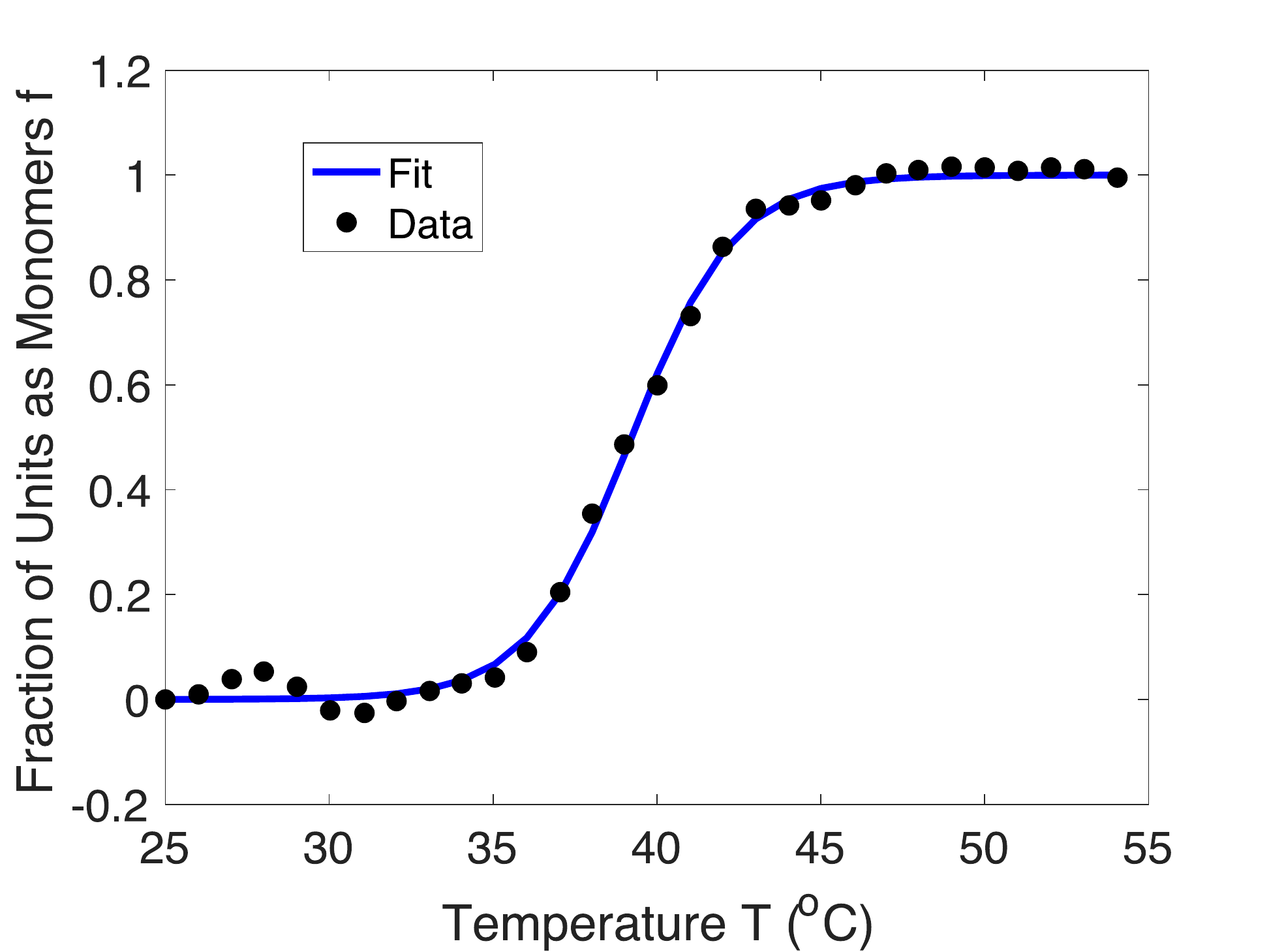}
	\caption{Fraction of TlpA units in the monomer state as a function of temperature. The black dots are the experimental data (open circles in Fig.\ 5C of Ref.\ [16] of the main text), while the blue line is the sigmoidal fit (Eq.\ \ref{eq:sigmoidFit}) with half-maximal temperature $T_M =39\ ^{\circ}$C and width $\Delta T= 6.3\ ^{\circ}$C.}
	\label{fig:hurmeFit}
\end{figure}

In their experiment, Hurme et al.\ used circular dichroism to infer the fraction of TlpA units in the monomeric state as a function of temperature \textit{in vitro} (Ref.\ [16] of the main text). The resulting curves are dependent on the supplied concentration of subunits, and they considered two concentrations: 0.12 $\mu$M and 3.61 $\mu$M. From a blotting analysis, they estimated the \textit{in vivo} concentration as 0.36 $\mu$M at 28 ${}^{\circ}$C and 0.6 $\mu$M at 37 ${}^{\circ}$C. Both of these are closer to the 0.12 $\mu$M value used \textit{in vitro}, so we use this case. We fit the fraction of TlpA units in the monomeric state as a function of temperature with a sigmoid
\begin{equation}\label{eq:sigmoidFit}
	f(T) = \frac{1}{1+\exp(-4(T-T_M)/\Delta T)}
\end{equation}
using the method of least squares to determine the parameters. We find $T_M =39\ ^{\circ}$C and $\Delta T= 6.3\ ^{\circ}$C. The fit is shown in Fig.\ \ref{fig:hurmeFit}.

\section{Derivation of Eqs.\ 3-5 of the Main Text}

\subsection{Properties of the Ornstein-Uhlenbeck Process}
The Ornstein-Uhlenbeck process is important because it appears as the linearization of any Markovian chemical master equation. The stochastic differential equation takes the form\begin{equation}
d\vec{X}_t = \mathcal{J} (\vec{X}_t-\vec{\mu}) dt + d\vec{N}_t,
\end{equation}
where $\vec{X}_t$ is $n$ dimensional, the Jacobian matrix $\mathcal{J}$ and the mean $\vec{\mu}$ are constant in time, and $\vec{N}_t$ is a vector of $n$ correlated and scaled Wiener processes where the mean is zero and the covariances are given by
\begin{equation}
\langle \vec{N}_t \vec{N}^T_s\rangle = \text{min}(s,t)\Sigma,
\end{equation}
for $\Sigma$ symmetric and positive-definite. The general solution is 
\begin{equation}\label{eq:OUSol}
\vec{X}_t = \vec{\mu} + e^{\mathcal{J} t}(\vec{X}_0-\vec{\mu}) + \int_{0}^{t} e^{\mathcal{J}(t-s)} d\vec{N}_s.
\end{equation}
The steady state mean is $\vec{\mu}$, and the steady state covariance matrix is computed through
\begin{equation}
\mathcal{C} = \lim\limits_{t\rightarrow\infty} \left\langle (\vec{X}_t-\vec{\mu}) (\vec{X}_t-\vec{\mu})^T \right\rangle.
\end{equation}
In order for this limit to exist, $\mathcal{J}$ must have eigenvalues with negative real parts, and we assume this to be the case, as this also implies that the deterministic system is stable. The It{\^o} isometry (Ref.\ [30] of the main text) can be used to simplify this to an integral
\begin{equation}\label{eq:OUCovarMat}
\mathcal{C} = \int_{0}^{\infty} e^{\mathcal{J}t}\Sigma e^{\mathcal{J}^T t}dt.
\end{equation}
By using integration by parts, we find the Lyapunov equation
\begin{equation}\label{eq:lyapunovEqn}
\mathcal{J}\mathcal{C} +\mathcal{C}\mathcal{J}^T +\Sigma =0.
\end{equation}
This is easier to solve in practice, since it is linear in the components of $\mathcal{C}$.

Now we compute the cross-correlation matrix. We start in steady state, so we assume that $\vec{X}_0$ is gaussian distributed with mean $\vec{\mu}$ and covariances given by the steady state covariance matrix $\mathcal{C}$. The cross-correlation matrix is defined through 
\begin{equation}
\mathcal{C}(\tau) = \left\langle (\vec{X}_{t+\tau}-\vec{\mu}) (\vec{X}_t-\vec{\mu})^T \right\rangle,
\end{equation} 
where stationarity removes the $t$-dependence. Progress can be made by using the It{\^o} isometry again and proceeding by cases depending on the sign of $\tau$. This yields the result
\begin{equation}\label{eq:OUCorr}
\mathcal{C}(\tau) = \begin{cases}
e^{\mathcal{J} \tau} \mathcal{C}, \quad \tau>0, \\
\mathcal{C}e^{-\mathcal{J}^T \tau}, \quad \tau<0.
\end{cases}
\end{equation}

Now let's put this all together. If $\vec{X}$ is a stationary stochastic process, the covariances of the time average $\mathcal{C}_{\text{TA}}$ over the window $[0,\tau]$ are related to the cross-correlations $\mathcal{C}(t)$ via
\begin{equation}
\mathcal{C}_{\text{TA}}(\tau) = \tau^{-2} \int_{0}^{\tau} \int_{0}^{\tau} \mathcal{C}(t-t')dt dt'. 
\end{equation}
As before, we change to the pair of variables $(\Delta, t')$, with $\Delta = t-t'$, and switch the order of integration so that $t'$ is integrated first. This leads to
\begin{equation}
\mathcal{C}_{\text{TA}}(\tau) = \tau^{-2}\left[ \int_{0}^{\tau} (\tau-\Delta) \mathcal{C}(\Delta)d\Delta + \int_{-\tau}^{0} (\tau+\Delta) \mathcal{C}(\Delta)d\Delta \right].
\end{equation}
Using the specific form of our cross-correlation matrix and integrating by parts gives
\begin{equation}
\mathcal{C}_{\text{TA}}(\tau) = \tau^{-2}\left[ -\tau \mathcal{J}^{-1}\mathcal{C} +\mathcal{J}^{-2}\left[e^{\mathcal{J} \tau} -\mathbb{I}\right]\mathcal{C} -\tau  \mathcal{C}(\mathcal{J}^T)^{-1} + \mathcal{C}(\mathcal{J}^T)^{-2}\left[e^{\mathcal{J}^T \tau} -\mathbb{I}\right] \right],
\end{equation}
the inverse of $\mathcal{J}$ exists since the eigenvalues have negative real parts and we have used the shorthand $\mathcal{J}^{-2}=(\mathcal{J}^{-1})^2$. We can simplify things a bit further by using Eq.\ \ref{eq:lyapunovEqn} to yield
\begin{equation}\label{eq:OUTACovar}
\mathcal{C}_{\text{TA}}(\tau) = \tau^{-2}\left[\tau  \mathcal{J}^{-1}\Sigma (\mathcal{J}^T)^{-1}+ \mathcal{J}^{-2}\left[e^{\mathcal{J} \tau} -\mathbb{I}\right]\mathcal{C}  + \mathcal{C}(\mathcal{J}^T)^{-2}\left[e^{\mathcal{J}^T \tau} -\mathbb{I}\right] \right].
\end{equation}
The first term is what we would find in the zero-frequency limit of the power spectrum.

\subsection{General Setting for the Biochemical Models}

In our model, we have a monomer that can reversibly form a dimer. The monomer is actively produced, but the dimer represses the production of the monomer, and both are lost via dilution, leading to the reactions
\begin{equation}
\begin{gathered}
\ce{2 M <=>[$k_d$][$k_m$] D}, \quad \ce{ M <=>[$k^-$][$g(d_t)$] \varnothing}, \quad \ce{ D ->[$k^-$] \varnothing}, \text{ with } g(d_t) = \frac{k^+}{1+\alpha d_t}.
\end{gathered}
\end{equation}

One can write the Kramers-Moyal expansion for the stochastic reactions. Doing so out to second-order derivatives yields a Fokker-Planck equation, which corresponds to the following stochastic differential equations
\begin{equation}\label{eq:KMSDEs}
\begin{gathered}
dd_t = [k_d m_t^2 - (k_m + k^-) d_t]dt +\sqrt{k_d m_t^2 +k_m d_t}dW^{(1)}_t+\sqrt{k^- d_t} dW^{(2)}_t, \\
dm_t = \left[\frac{k^+}{1+\alpha d_t} +2k_m d_t - k^- m_t -2k_d m_t^2\right]dt -2\sqrt{k_d m_t^2 +k_m d_t}dW^{(1)}_t +\sqrt{\frac{k^+}{1+\alpha d_t} + k^- m_t} dW^{(3)}_t,
\end{gathered}
\end{equation}
where $W^{(1)}_t$, $W^{(2)}_t$, $W^{(3)}_t$ are standard, independent Wiener processes with variance $t$.

From the deterministic equations, it is easy to see that there is a unique positive steady state.
The fraction of TlpA molecules in the monomer state
\begin{equation}\label{eq:frac}
f = \frac{\overline{m}}{\overline{m} + 2\overline{d}},
\end{equation}
has been well characterized experimentally, where the bars denote the deterministic steady state or mean values. We can use this to solve for the dimer number in terms of the monomer number and fraction
\begin{equation}\label{eq:DforMF}
	\overline{d} = \frac{\overline{m}}{2} \left(\frac{1-f}{f}\right).
\end{equation}
Using the steady state condition from the dimer's equation of motion gives
\begin{equation}
\frac{k_m+k^-}{k_d} = \frac{\overline{m}^2}{\overline{d}}.
\end{equation}
We can use Eq.\ \ref{eq:DforMF} to eliminate the dimer, which gives
\begin{equation}\label{eq:KDForF}
k_m = \frac{2\overline{m}fk_d}{1-f}-k^-.
\end{equation}
We will make this substitution when solving the rate equations for $\overline{m}$.

\subsection{Derivation of Eq.\ 3 (Fixed Pool)}

For the fixed pool, we take $k^+$ and $k^-$ to zero. This makes $m_t+2d_t=n$ a conserved quantity that we call the pool size. We can eliminate the dimer from the dynamics
\begin{equation}
dm_t = [k_m (n-m_t) -2k_d m_t^2]dt-2\sqrt{k_d m_t^2 +k_m \left(\frac{n-m_t}{2}\right)}dW^{(1)}_t.
\end{equation}
We start by finding the steady state mean. We do so by identifying the $m$ value that causes the deterministic term to vanish. Using Eq.\ \ref{eq:KDForF}, this gives $\overline{m}=nf$, as expected. To convert noise in molecules to noise in a temperature estimate, we also need a linearization factor $d\overline{m}/dT$, which is just $nf'$.

Now we will linearize the system to find the fluctuations. In doing so, we will also confirm that the fixed point is linearly stable. Letting $\delta m_t = m_t-\overline{m}$ and expanding the equation to first-order in $\delta m$ in the deterministic term and zeroth-order in the noise term, we find
\begin{equation}
d(\delta m_t) =-2 k_d n\frac{f(2-f)}{1-f}\delta m_t dt-2\sqrt{2k_d f^2 n^2}dW^{(1)}_t
\end{equation}
The coefficient of the linearized deterministic term is negative, so the fixed point is deterministically stable. We see that the Jacobian and noise covariance matrix are
\begin{equation}\label{eq:FPJB}
\mathcal{J} = -2k_d n\frac{f(2-f)}{1-f}, \quad \Sigma = 8 k_d f^2 n^2.
\end{equation}
With these, we can find the variance in the time average or the maximum likelihood estimate of the mean, as the Lyapunov equation is trivial to solve for scalars. The steady state variance may be computed from Eq.\ \ref{eq:lyapunovEqn}
\begin{equation}\label{eq:FPVar}
	\sigma^2(m) = \frac{2 (1-f) f n}{2-f}.
\end{equation}
The variance in the time averaged monomer number may be computed from Eq.\ \ref{eq:OUTACovar} by using $\mathcal{C} = \sigma^2(m)$ and the expressions for $\mathcal{J}$ and $\Sigma$ in Eq.\ \ref{eq:FPJB}. From Eq.\ \ref{eq:OUCorr}, we see that the correlation timescale is $\tau_d = -1/\mathcal{J}$. This completes the derivation of Eq.\ 3 of the main text.

\subsection{Derivation of Eqs.\ 4 and 5 (Production-Dilution, without and with Feedback)}

\subsubsection{Deterministic Analysis}
The deterministic equations for the system are
\begin{equation}
\begin{gathered}
\dot{d} = k_d m^2 - (k_m+k^-) d, \\
\dot{m} = \frac{k^+}{1+\alpha d} +2k_m d - k^- m -2k_d m^2.
\end{gathered}
\end{equation}
The mean dimer number can be found from its equation of motion: $\overline{d} = k_d \overline{m}^2/(k_m+k^-).$ Using the dimer steady state equation and Eq.\ \ref{eq:DforMF}, we find that the monomer steady state value satisfies
\begin{equation}
0 =  \frac{k^+}{1+\alpha \overline{m}(1-f)/(2f)} - \frac{k^- \overline{m}}{f}.
\end{equation}
The production term decreases from $k^+$ to $0$ monotonically for $\overline{m}>0$, while the loss term increases monotonically from $0$ to infinity, so there is exactly one stable, positive fixed point. We find that the positive root is
\begin{equation}\label{eq:PDFMeanFull}
\overline{m} = \frac{f}{\alpha(1-f)}\left[\sqrt{1+\frac{2(1-f)\alpha k^+}{k^-}} -1\right],
\end{equation}
as in Eq.\ 5 of the main text. It will be helpful to compute the mean monomer number in the absence of autorepression. This is done by taking $\alpha\rightarrow 0$, where we find
\begin{equation}\label{eq:m0}
	m_0 := \lim\limits_{\alpha\rightarrow 0} \overline{m} = \frac{fk^+}{k^-}.
\end{equation}
as given above Eq.\ 4 of the main text. We can treat $m_0$ as a free parameter and solve for $f$ in terms of $\alpha$, $m_0$, and $\overline{m}$
\begin{equation}\label{eq:fForMStar}
	f = \frac{\overline{m}^2}{\overline{m}^2+2 \alpha^{-1} (m_0-\overline{m})}.
\end{equation}
This will be useful in simplifying expressions later on.

We show that the fixed point is stable. The Jacobian at the fixed point is
\begin{equation}\label{eq:simpJac}
	\mathcal{J}= \begin{bmatrix}
		-k_m-k^- & 2k_d \overline{m} \\
		2k_m - \dfrac{\alpha k^+}{(1+\alpha \overline{d})^2} & -k^- -4k_d \overline{m}
	\end{bmatrix}.
\end{equation}
The eigenvalues of this matrix both have negative real parts if the trace is negative and the determinant is positive. Since $\overline{m}>0$, it is trivial to see that this has a negative trace. Using Eqs.\ \ref{eq:DforMF} and \ref{eq:KDForF}, it follows that the determinant is positive, so we conclude that this fixed point is stable.

\subsubsection{Stochastic Analysis}

Linearizing the noise term, we can read off the form of $\vec{N}_t$
\begin{equation}\label{eq:NNoise}
	\begin{bmatrix}
		N^{(1)}_t \\
		N^{(2)}_t
	\end{bmatrix} = 
	\begin{bmatrix}
		\sqrt{k_d \overline{m}^2 +k_m \overline{d}}W^{(1)}_t+\sqrt{k^- \overline{d}} W^{(2)}_t \\
		-2\sqrt{k_d \overline{m}^2 +k_m \overline{d}}W^{(1)}_t +\sqrt{\dfrac{k^+}{1+\alpha \overline{d}} + k^- \overline{m}} W^{(3)}_t
	\end{bmatrix}.
\end{equation}
We just need to identify the matrix $\Sigma$, which may be readily computed from the previous expression and simplified using the steady state equations and the expression for dimer from Eq.\ \ref{eq:DforMF}
\begin{equation}\label{eq:PDFSigma}
	\begin{aligned}
		\Sigma &= \begin{bmatrix}
			k_d \overline{m}^2 +k_m \overline{d} +k^- \overline{d} & -2\left(k_d \overline{m}^2 +k_m \overline{d}\right) \\
			-2\left(k_d \overline{m}^2 +k_m \overline{d}\right) & 4\left(k_d \overline{m}^2 +k_m \overline{d}\right) +\dfrac{k^+}{1+\alpha \overline{d}} + k^- \overline{m}
		\end{bmatrix} \\
	&= \begin{bmatrix}
		2k_d \overline{m}^2 & -2\left(2k_d \overline{m}^2 -k^- \dfrac{\overline{m}(1-f)}{2f}\right) \\
		-2\left(2k_d \overline{m}^2 -k^- \dfrac{\overline{m}(1-f)}{2f}\right) & 8k_d \overline{m}^2 + k^- \dfrac{\overline{m}(5f-1)}{2f}
	\end{bmatrix}.
	\end{aligned}
\end{equation}
With this, our system of SDEs is in the canonical form for an OU process. The covariance matrix may be determined from Eq.\ \ref{eq:lyapunovEqn}. Using Eq.\ \ref{eq:fForMStar} to eliminate the fraction, we find that,
\begin{equation}\label{eq:PDFCovar}
\begin{split}
\mathcal{C}_{1,1}&=\frac{\left(\overline{m}-m_0\right) k_d \left(-2 \alpha  \overline{m}^4 \left(\alpha  \overline{m}-2\right)+m_0 \overline{m}^2 \left(\alpha  \overline{m} \left(3 \alpha 
	\overline{m}-16\right)+14\right)+4 m_0^2 \overline{m} \left(3 \alpha  \overline{m}-7\right)+14 m_0^3\right)}{\alpha  \left(2 \overline{m}^2 \left(\alpha 
	\overline{m}-2\right)-3 m_0 \overline{m} \left(\alpha  \overline{m}-4\right)-8 m_0^2\right) \left(\overline{m} k_d \left(\overline{m} \left(\alpha  \overline{m}-4\right)+4
	m_0\right)+k^- \left(m_0-\overline{m}\right)\right)}\\
	&-\frac{k^- m_0 \left(\overline{m}-m_0\right){}^2 \left(\overline{m} \left(\alpha  \overline{m}-4\right)+4 m_0\right)}{\alpha  \overline{m} \left(2 \overline{m}^2
		\left(\alpha  \overline{m}-2\right)-3 m_0 \overline{m} \left(\alpha  \overline{m}-4\right)-8 m_0^2\right) \left(\overline{m} k_d \left(\overline{m} \left(\alpha 
		\overline{m}-4\right)+4 m_0\right)+k^- \left(m_0-\overline{m}\right)\right)},\\
\mathcal{C}_{1,2} &=\mathcal{C}_{2,1}=\frac{\overline{m} \left(\overline{m}-m_0\right){}^2 \left(\overline{m} k_d \left(4 \overline{m} \left(\alpha  \overline{m}-2\right)+9 m_0\right)+k^- \left(\overline{m}
	\left(\alpha  \overline{m}-2\right)+2 m_0\right)\right)}{\left(2 \overline{m}^2 \left(\alpha  \overline{m}-2\right)-3 m_0 \overline{m} \left(\alpha 
	\overline{m}-4\right)-8 m_0^2\right) \left(\overline{m} k_d \left(\overline{m} \left(\alpha  \overline{m}-4\right)+4 m_0\right)+k^-
	\left(m_0-\overline{m}\right)\right)}, \\
\mathcal{C}_{2,2} &= -\frac{\overline{m}^2 k_d \left(16 \overline{m}^3 \left(\alpha  \overline{m}-2\right)+m_0 \overline{m}^2 \left(\alpha  \overline{m} \left(2 \alpha 
	\overline{m}-47\right)+128\right)+m_0^2 \overline{m} \left(31 \alpha  \overline{m}-160\right)+64 m_0^3\right)}{2 \left(2 \overline{m}^2 \left(\alpha 
	\overline{m}-2\right)-3 m_0 \overline{m} \left(\alpha  \overline{m}-4\right)-8 m_0^2\right) \left(\overline{m} k_d \left(\overline{m} \left(\alpha  \overline{m}-4\right)+4
	m_0\right)+k^- \left(m_0-\overline{m}\right)\right)} \\
&+ \frac{k^- \overline{m}^2 \left(\overline{m}-m_0\right) \left(\alpha  \overline{m}-2\right) \left(\alpha  \overline{m} \left(\overline{m} \left(\alpha 
	\overline{m}-2\right)+m_0 \left(10-\alpha  \overline{m}\right)\right)+m_0\right)}{\alpha  m_0 \left(-2 \overline{m}^2 \left(\alpha  \overline{m}-2\right)+3 m_0
	\overline{m} \left(\alpha  \overline{m}-4\right)+8 m_0^2\right) \left(\overline{m} k_d \left(\overline{m} \left(\alpha  \overline{m}-4\right)+4 m_0\right)+k^-
	\left(m_0-\overline{m}\right)\right)} \\
&+ \frac{k^- m_0 \left(\overline{m}-m_0\right) \left(\overline{m}^2 \left(\alpha  \overline{m} \left(18 \alpha  \overline{m}-59\right)-16\right)+m_0 \overline{m} \left(29
	\alpha  \overline{m}+20\right)-8 m_0^2\right)}{2 \alpha  \overline{m} \left(2 \overline{m}^2 \left(\alpha  \overline{m}-2\right)-3 m_0 \overline{m} \left(\alpha 
	\overline{m}-4\right)-8 m_0^2\right) \left(\overline{m} k_d \left(\overline{m} \left(\alpha  \overline{m}-4\right)+4 m_0\right)+k^-
	\left(m_0-\overline{m}\right)\right)}.
\end{split}
\end{equation}
As before, the variance in the time averaged monomer number may be computed from Eq.\ \ref{eq:OUTACovar} by using this covariance matrix with the expressions for $\mathcal{J}$ and $\Sigma$ in Eqs.\ \ref{eq:simpJac} and \ref{eq:PDFSigma} respectively.
The variance in the pool size may be computed from these results according to
\begin{equation}
	\sigma^2(n) = \text{cov}(2d+m,2d+m) = 4\mathcal{C}_{1,1} + 4\mathcal{C}_{1,2} +\mathcal{C}_{2,2},
\end{equation}
where ``cov" denotes the covariance.

We now describe how to arrive at Eq.\ 4 of the main text. In the limit that protein loss is much slower than dimerization $k^-\ll k_d \overline{m}$, we find that the variance $\sigma^2(m) = \mathcal{C}_{2,2}$ and $\sigma^2(n)$ simplify to
\begin{equation}
	\begin{gathered}
		\sigma^2_{\rm PDF}(m)=-\frac{2 \alpha ^2 m_0 \overline{m}^5+\alpha  \left(16 \overline{m}-31 m_0\right) \left(\overline{m}-m_0\right) \overline{m}^3-32 \left(\overline{m}-2 m_0\right)
			\left(\overline{m}-m_0\right){}^2 \overline{m}}{2 \left(\alpha  \overline{m}^2 \left(2 \overline{m}-3 m_0\right)-4 \left(\overline{m}-2 m_0\right)
			\left(\overline{m}-m_0\right)\right) \left(\overline{m} \left(\alpha  \overline{m}-4\right)+4 m_0\right)}, \\
		\sigma^2_{\rm PDF}(n)= -\frac{m_0 \left(\overline{m} \left(\alpha  \overline{m}-4\right)+4 m_0\right) \left(\overline{m} \left(2 \alpha  \overline{m}-7\right)+7 m_0\right)}{2 \alpha 
			\overline{m} \left(\alpha  \overline{m}^2 \left(2 \overline{m}-3 m_0\right)-4 \left(\overline{m}-2 m_0\right) \left(\overline{m}-m_0\right)\right)}.
	\end{gathered}
\end{equation}
In the limit of no feedback, $\alpha\rightarrow 0$, the expressions simplify further to
\begin{equation}
	\sigma^2_{\text{PD}}(m) =\frac{f \left(5 f^2-17 f+16\right) k^+}{4 (f-2)^2 k^-}, \quad \sigma^2_{\text{PD}}(n) = \frac{(7-3 f) k^+}{4 k^-}.
\end{equation}
We compute $\sigma^2_{\text{FP}}(m)$ according to Eq.\ \ref{eq:FPVar} but take $n=\overline{m}/f$, where the expression for $\overline{m}$ is taken from Eq.\ \ref{eq:PDFMeanFull}. Eq.\ 4 in the main text may be verified by computing $(\sigma_{\text{PDF/PD}}^2(m)-\sigma^2_{\text{FP}}(m))/\sigma^2_{\text{PDF/PD}}(n)$, which simplifies to $f^2/(2-f)^2$ for both nonzero $\alpha$ (PDF) and zero $\alpha$ (PD).

\subsection{Optimal Autorepression}\label{sec:SetValues}
The autorepression strength $\alpha$ has competing effects such that there is an optimal strength that minimizes the relative error.
When the timescale separation is large, $k^- \ll k_d\overline{m}$, we find that the relative error $\sigma(m)/|d\overline{m}/dT| = \mathcal{C}_{2,2}^{1/2}/|d\overline{m}/dT|$ only depends on $\alpha$, $f$, and the ratio $k^+/k^-$. To find the optimal $\alpha$, we set $f=1/2$, find where the derivative of the relative error with respect to $\alpha$ vanishes, and check that the second derivative with respect to $\alpha$ at that point is positive. There is exactly one positive $\alpha$ value where the derivative vanishes, and it occurs at
\begin{equation}
\label{eq:optalpha}
	\alpha = \alpha^* = \frac{1.75}{k^+/k^-},
\end{equation}
as stated in the main text. The second derivative at this point is $0.0105\times (k^+/k^-)^{3/2}$, so this is a local minimum. We find that the relative error is $\mathcal{O}(\alpha^{1/4})$ as $\alpha\rightarrow\infty$ in both cases, so this is the global minimum. Using the ratio $k^+/k^- =2521$ estimated from experiments, this leads to $\alpha^* =6.9\times 10^{-4}$.

\section{Monomer Readout vs.\ Dimer Readout}

\begin{figure}[h]
	\centering
	\includegraphics[width=0.5\columnwidth]{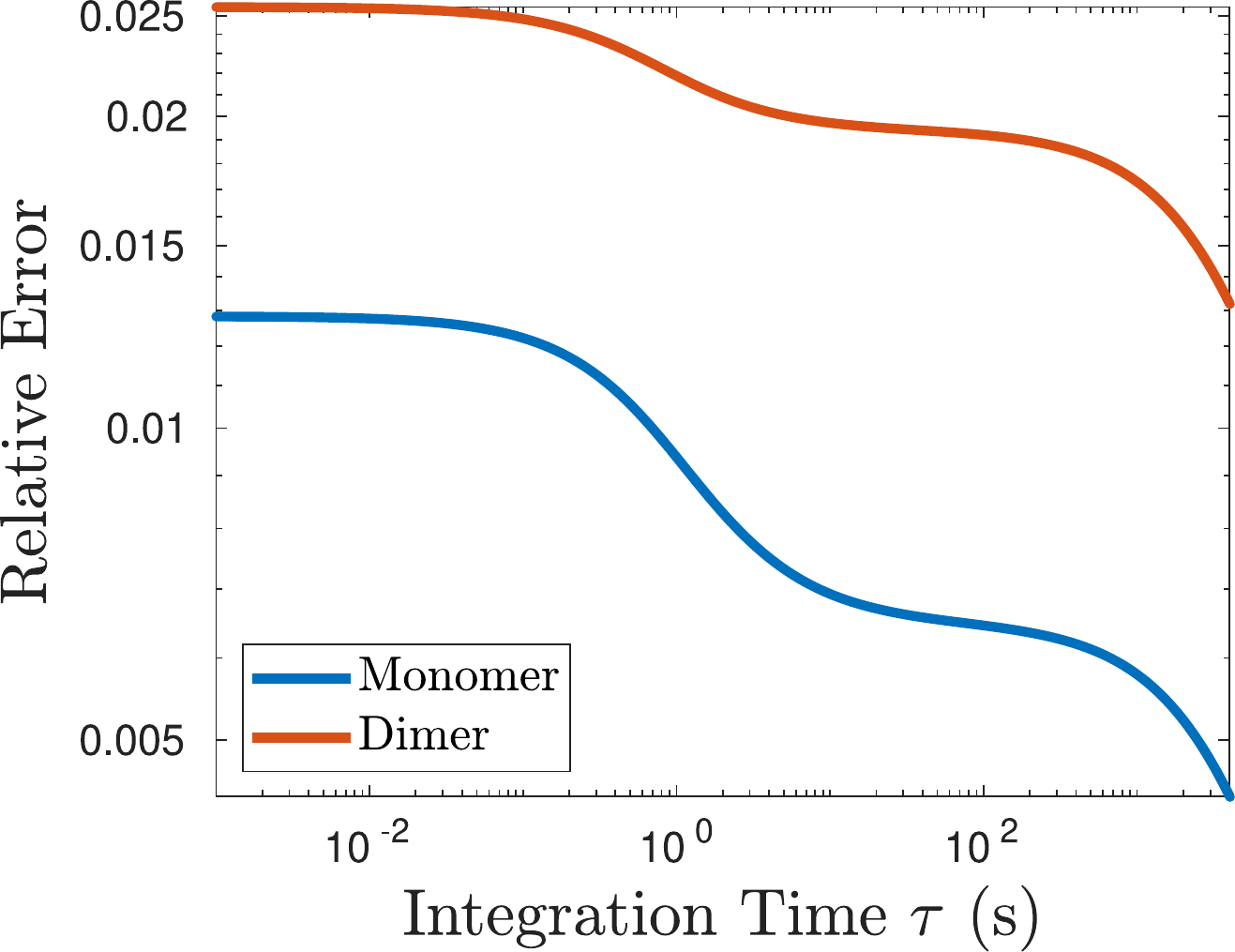}
	\caption{Comparison of the relative error in temperature sensing inferred from the monomer number vs.\ from the dimer number. Parameters are as in the main text: $f=1/2$, $\alpha= 6.94\times 10^{-4}$, $k_d=7.1\times 10^{-4}$ $\text{s}^{-1}$, $k^-= 5.5\times 10^{-4}$ $\text{s}^{-1}$, and $k^+/k^-=2521$.}
	\label{fig:dimer}
\end{figure}

Here we compare the temperature estimation error inferred from the monomer number (as in the main text) and from the dimer number. We see in Fig.\ \ref{fig:dimer} that qualitatively the error has a similar dependence on the integration time in the two cases, and quantitatively the two only differ by a factor of order unity (ranging from approximately two to three depending on the integration time).

\section{Maximum Likelihood Estimation}
\subsection{Trajectory Probability}

We want to incorporate all of the available information to derive the limits to cellular performance. This requires the probability of observing a specified trajectory. Suppose that we have a stochastic differential equation with additive noise
\begin{equation}
	d\vec{X}_t = \vec{F}(\vec{X}_t,t)dt + d\vec{N}_t,
\end{equation}
where $\vec{X}_t$ is $n$-dimensional, and $\vec{N}$ is an $n$-dimensional scaled Wiener process where the mean is zero and the covariances are given by
\begin{equation}
	\langle \vec{N}_t \vec{N}^T_s \rangle = \text{min}(s,t)\Sigma,
\end{equation}
for $\Sigma$ symmetric and positive definite. We define the current at time $t$ to be
\begin{equation}
\vec{J}(\vec{X}_t,t) = \vec{F}(\vec{X}_t,t)dt + d\vec{N}_t.
\end{equation}
We can formally write the probability density of observing a trajectory by discretely sampling it at $M$ points in time separated by time $\Delta t$ and computing
\begin{equation}
P(\{\vec{X}_{t}\}_{t>0}|\vec{X}_0) = \lim\limits_{M\rightarrow \infty} \left\langle \prod\limits_{j=1}^{M} \delta(\vec{X}_{j} - \vec{X}_{j-1}-\vec{J}_{j-1})\right\rangle_J,
\end{equation}
where we used the It{\^o} discretization, take $\Delta t\rightarrow 0$ such that $M\Delta t =\tau$ is constant, and use a subscript $j$ to indicate evaluation at $j\Delta t$ and $\vec{X}_{j\Delta t}$. Since we specify the trajectory we are interested in, $\vec{J}_j$ is a gaussian random variable that is linearly related to the scaled Wiener processes via
\begin{equation}
\Delta\vec{N}_j = \vec{J}_j -\vec{F}_j \Delta t.
\end{equation}
The increments of the Wiener process at different times are independent, while at the same time their covariance is
\begin{equation}
\left\langle \Delta\vec{N}_j \Delta\vec{N}_j^T \right\rangle = (\Delta t) \Sigma.
\end{equation}
It follows that the jump distribution at one instant is
\begin{equation}
P(\vec{J}_j) = P(\Delta\vec{N}_j) \left| \det\left(\frac{\partial (\Delta \vec{N}_j)}{\partial \vec{J}_j}\right)\right|= \frac{\exp\left(-\frac{1}{2\Delta t} \langle \vec{J}_j -\vec{F}_j \Delta t, \Sigma^{-1}(\vec{J}_j -\vec{F}_j \Delta t)\rangle  \right)}{\sqrt{(2\pi)^{n} \det(\Sigma)}}.
\end{equation}
Since the process is Markovian, we may evaluate the average of the product of deltas term-by-term
\begin{equation}
P(\{\vec{X}_{t}\}_{t>0}|\vec{X}_0) = \lim\limits_{M\rightarrow \infty} \prod\limits_{j=1}^{M}\left\langle  \delta(\vec{X}_{j} - \vec{X}_{j-1}-\vec{J}_{j-1})\right\rangle_J.
\end{equation}
Using the delta functions leads to the result
\begin{equation}
P(\{\vec{X}_{t}\}_{t>0}|\vec{X}_0) = \lim\limits_{M\rightarrow \infty} \prod\limits_{j=1}^{M}\frac{\exp\left(-\frac{1}{2\Delta t} \langle \vec{X}_{j} - \vec{X}_{j-1} -\vec{F}_j \Delta t, \Sigma^{-1}(\vec{X}_{j} - \vec{X}_{j-1} -\vec{F}_j \Delta t)\rangle \right)}{\sqrt{(2\pi)^{n} \det(\Sigma)}}.
\end{equation}
This is the well-known Onsager-Machlup functional. To get the probability of the full trajectory, we need to weight this by the probability of a given initial condition
\begin{equation}
\begin{aligned}
P(\{\vec{X}_{t}\}_{t\geq 0}) &= P(\vec{X}_0) \lim\limits_{M\rightarrow \infty} \prod\limits_{j=1}^{M}\frac{\exp\left(-\frac{1}{2\Delta t} \langle \vec{X}_{j} - \vec{X}_{j-1} -\vec{F}_j \Delta t, \Sigma^{-1}(\vec{X}_{j} - \vec{X}_{j-1} -\vec{F}_j \Delta t)\rangle \right)}{\sqrt{(2\pi)^{n} \det(\Sigma)}}.
\end{aligned}
\end{equation}

\subsection{The Maximum Likelihood Estimate}

For the case of an Ornstein-Uhlenbeck process, we have
\begin{equation}
\vec{F}_j = \mathcal{J}(\vec{X}_j-\vec{\mu}).
\end{equation}
and the steady state distribution
\begin{equation}
P(\vec{X}_0)=\frac{\exp\left(-\frac{1}{2} \left\langle \vec{X}_0 -\vec\mu, \mathcal{C}^{-1}(\vec{X}_0-\vec{\mu})\right\rangle \right)}{\sqrt{(2\pi)^{n} \det(\mathcal{C})}}.
\end{equation}
We can re-write our functional as
\begin{equation}
\begin{aligned}
P(\vec{X}_{t\geq 0}) &\sim \frac{\exp\left(-\frac{1}{2} \left\langle \vec{X}_0 -\vec\mu, \mathcal{C}^{-1}(\vec{X}_0-\vec{\mu})\right\rangle \right)}{\sqrt{(2\pi)^{n} \det(\mathcal{C})}} \\
&\times \prod\limits_{j=1}^{M}\frac{\exp\left(-\frac{1}{2\Delta t} \langle \vec{X}_{j} - \vec{X}_{j-1} -\vec{F}_j \Delta t, \Sigma^{-1}(\vec{X}_{j} - \vec{X}_{j-1} -\vec{F}_j \Delta t)\rangle \right)}{\sqrt{(2\pi)^{n} \det(\Sigma)}},
\end{aligned}
\end{equation}
where we are working with the expression at finite $M$ and then taking the $M\rightarrow\infty$ limit. Generally, the maximum likelihood estimate is computed via
\begin{equation}
\hat{\vec{\theta}} = \text{argmax}_{\vec{\theta}} P(\vec{x}|\vec{\theta}),
\end{equation}
where $\vec{x}$ is the data and $\vec{\theta}$ are the parameters influencing the distribution (Ref.\ [27] of the main text). Since the logarithm is monotone, this maximization is often computed by taking the logarithm and differentiating. This gives
\begin{equation}
\nabla_{\vec{\mu}}\log P \sim -\mathcal{C}^{-1}(\vec{\mu}-\vec{X}_0) - \mathcal{J}^T \Sigma^{-1} \sum\limits_{j=1}^{M}(\vec{X}_{j} - \vec{X}_{j-1} +\mathcal{J}(\vec{\mu}-\vec{X}_{j-1}) \Delta t).
\end{equation}

Let's simplify this expression. The difference of $X$ terms form a telescoping sum that simplifies to
\begin{equation}
\sum\limits_{j=1}^{M}(\vec{X}_{j} - \vec{X}_{j-1}) = \vec{X}_{\tau} - \vec{X}_0,
\end{equation}
using the fact that $M\Delta t=\tau$. The $\mu$ term from the dynamics simplifies
\begin{equation}
\sum\limits_{j=1}^{M}\mathcal{J}\vec{\mu} \Delta t = \mathcal{J} \vec{\mu} \tau.
\end{equation}
The last term from the dynamics is the definition of an It{\^o} integral
\begin{equation}
-\sum\limits_{j=1}^{M}\mathcal{J}\vec{X}_{j-1} \Delta t = -\mathcal{J} \int_0^{\tau}\vec{X}_t dt.
\end{equation}
This leads to the simplified expression
\begin{equation}
\nabla_{\vec{\mu}}\log P = -\mathcal{C}^{-1}(\vec{\mu}-\vec{X}_0) - \mathcal{J}^T \Sigma^{-1} \left[\vec{X}_{\tau} - \vec{X}_{0} +\mathcal{J}\vec{\mu}\tau-\mathcal{J}\int_{0}^{\tau}\vec{X}_{t}dt \right].
\end{equation}
Our maximum likelihood estimate is determined by finding the $\vec{\mu}$ that causes the gradient to vanish, and this is
\begin{equation}
\begin{aligned}
\hat{\vec{\mu}} = \left(\mathcal{C}^{-1} +\mathcal{J}^T \Sigma^{-1} \mathcal{J}\tau\right)^{-1} \left[\left(\mathcal{C}^{-1} +\mathcal{J}^T \Sigma^{-1} \right)\vec{X}_0 - \mathcal{J}^T \Sigma^{-1} \vec{X}_{\tau} +\mathcal{J}^T \Sigma^{-1}\mathcal{J}\int_{0}^{\tau}\vec{X}_{t}dt\right].
\end{aligned}
\end{equation}
Note that this is an unbiased estimator, as the mean of $\vec{X}_t$ is $\vec{\mu}$ for all $t$ for our given initial condition. Furthermore, since it is a linear combination of gaussian random variables, it is also a gaussian random variable. In the long time limit, the time average term dominates. This can be written as the uniform time average plus corrections that vanish at long times. It is straightforward, albeit tediuous, to compute the covariance matrix. After a lot of cancellation, we find
\begin{equation}
\mathcal{C}_{\vec{\mu},\vec{\mu}} = \left\langle (\hat{\vec{\mu}}-\vec{\mu})(\hat{\vec{\mu}}-\vec{\mu})^T\right\rangle = \left(\mathcal{C}^{-1} +\mathcal{J}^T\Sigma^{-1} \mathcal{J}\tau\right)^{-1},
\end{equation}
This also has the same short- and long-time asymptotics as the naive time average.

\subsection{Cramer-Rao Bound}

The Cramer-Rao bound specifies the best that an estimate can possibly do (Ref.\ [27] of the main text). The multivariate Cramer-Rao bound states that, for an unbiased estimator,
\begin{equation}
\mathcal{C}_{\vec{\theta},\vec{\theta}} \geq \mathcal{I}(\vec{\theta})^{-1},
\end{equation}
where $\mathcal{I}(\vec{\theta})$ is the Fisher information
\begin{equation}\label{eq:fish}
\mathcal{I}_{i,j}(\vec{\theta}) := \left\langle \left(\frac{\partial \log P(\vec{x}|\vec{\theta})}{\partial\theta_i}\right)\left(\frac{\partial \log P(\vec{x}|\vec{\theta})}{\partial\theta_j}\right) \right\rangle = -\left\langle \frac{\partial^2 \log P(\vec{x}|\vec{\theta})}{\partial\theta_i \partial \theta_j} \right\rangle,
\end{equation}
where the last step assumes differentiability and uses integration by parts. Here $A\geq B$ means that $A-B$ is a positive semi-definite matrix. Note that any positive semi-definite matrix $M$ has $\vec{v}^T M \vec{v}\geq 0$ for any real vector $\vec{v}$. Taking $\vec{v}$ to be any standard unit vector gives the inequality
\begin{equation}
\left[\mathcal{C}_{\vec{\theta},\vec{\theta}}\right]_{i,i} \geq \left[ \mathcal{I}(\vec{\theta})^{-1} \right]_{i,i}.
\end{equation}
The variances of the estimates are bounded below by the diagonal elements of the inverse Fisher matrix. When the matrix ``inequality" is saturated, the variances of the estimates are minimized.

Focusing on a particular component of the mean, we have
\begin{equation}
\frac{\partial \log P}{\partial \mu_i} = -\sum_k\mathcal{C}^{-1}_{i,k}\mu_k- \sum_k (\mathcal{J}^T \Sigma^{-1} \mathcal{J})_{i,k}\mu_k\tau +f_i(\vec{X}),
\end{equation}
where $f_i(\vec{X})$ contains the $\vec{\mu}$-independent terms. Taking another derivative with respect to $\mu_j$ gives
\begin{equation}
\frac{\partial^2 \log P}{\partial \mu_i\partial \mu_j} = -\mathcal{C}^{-1}_{i,j}-  (\mathcal{J}^T \Sigma^{-1} \mathcal{J})_{i,j}\tau.
\end{equation}
This is a constant, so it is minus the Fisher information, see Eq.\ \ref{eq:fish}. We see that the inverse Fisher information matrix is equal to the covariance matrix $\mathcal{C}_{\vec{\mu},\vec{\mu}}$, so maximum likelihood estimation attains the minimum possible error.

\subsection{Comparison of Time Averaging and Maximum Likelihood}
Using the expressions derived above, we can compute the relative error using the maximum likelihood approach and compare it to the time average. This is shown in Fig.\ S3, where the parameter values are those used in the main text, and we see that the two curves are almost identical.
Although maximum likelihood is optimal, it assumes that the cell ``knows" the matrices $\mathcal{J}$ and $\Sigma$, but this result shows that the cell can come very close to the fundamental limit using naive time averaging.

\begin{figure}[h]
	\centering\includegraphics[width=0.5\columnwidth]{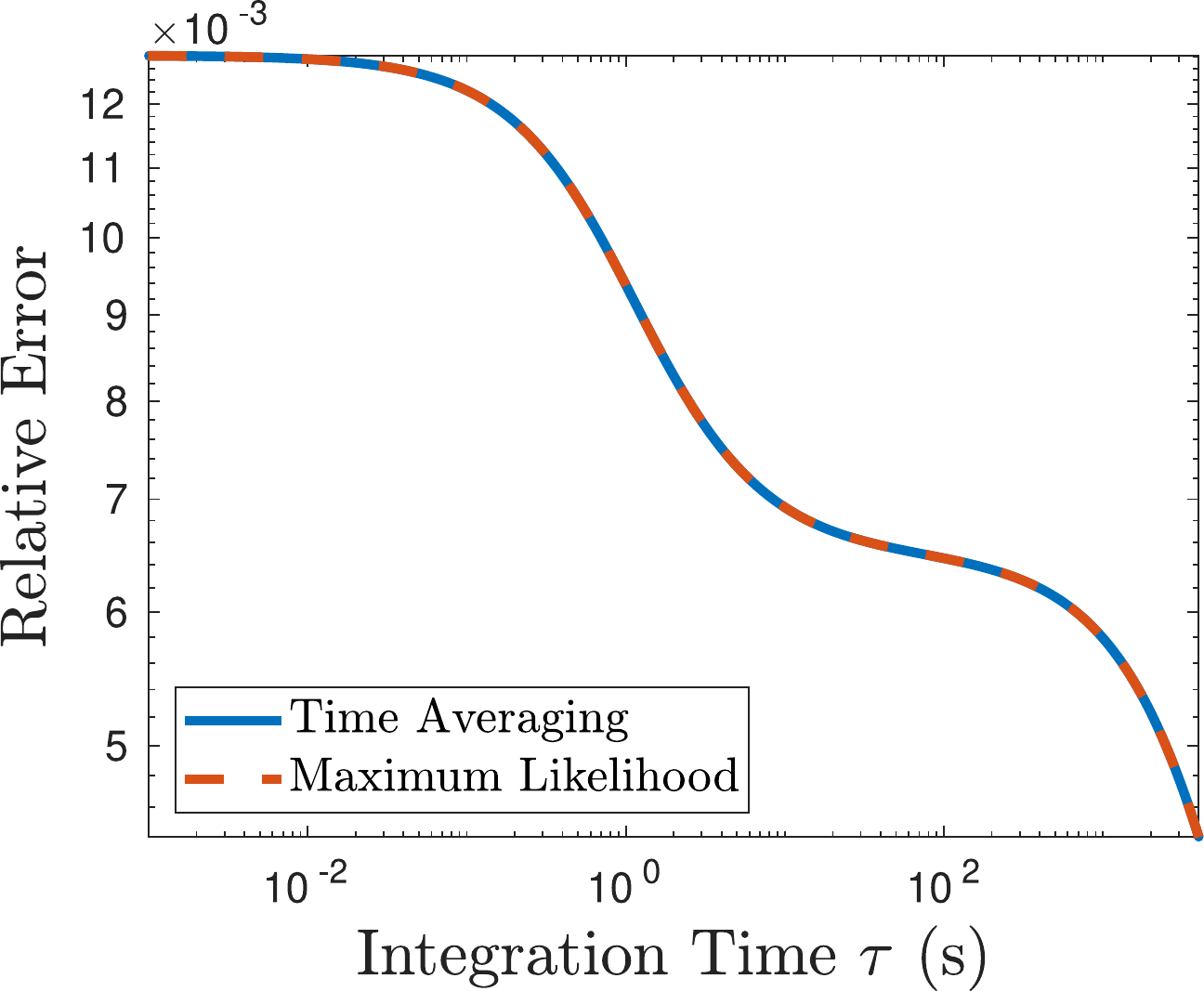}\label{fig:MLEvsTA}
	\caption{Comparison of the relative error in temperature sensing through maximum likelihood (red) and time averaging (blue). Parameter values match those used in the main text: $f=1/2$, $\alpha= 6.94\times 10^{-4}$, $k_d=7.1\times 10^{-4}$ $\text{s}^{-1}$, $k^-= 5.5\times 10^{-4}$ $\text{s}^{-1}$, and $k^+/k^-=2521$.}
\end{figure}

\section{Temperature Dependence of $k^+$, $k^-$, and $\alpha$}

In the main text, we account for the temperature dependence of the binding rates $k_d$ and $k_m$ via the experimentally characterized monomer fraction $f(T)$. We assume that the remaining parameters $k^+$, $k^-$, and $\alpha$ are temperature independent, but in principle they could depend on temperature. Therefore, we investigate the temperature dependence of these parameters here.
We will argue that $dk^+/dT$ and $d\alpha/dT$ may neglected based on experimental evidence and that $dk^-/dT$ as estimated from experiments is negligible quantitatively compared to $df/dT$.

We will begin by considering the experimental results regarding $k^+$ and $\alpha$. Hurme et al.\ performed a Miller assay on mutants where the region of gene coding for TlpA was removed and replaced with a reporter, while the promoter was unchanged (Ref.\ [16] of the main text). This assay reports the number of times that a gene has been expressed within a period of time, and it was not found to vary significantly with temperature in the mutants. Since $k^+$ is the production rate in the absence of autorepression (achieved here since the cells don't contain TlpA), we take this to mean that we may safely neglect its temperature dependence. This same evidence raises the possibility that the structure of the promoter does not radically change with temperature, which suggests that we may regard $\alpha$ as relatively insensitive to temperature. However, making a statement about $\alpha$ requires the dimer to be present. Hurme et al.\ also did further experiments without excising the coding region of the gene. The approach was to induce modifications to the TlpA binding site on the promoter similar to what would be seen upon induction to high temperature while leaving the temperature and the fraction $f$ constant. They tested the effects of DNA supercoiling via H-NS mutants and topoisomerase I and found that supercoiling did not contribute to derepression. They also applied ethanol stress, which is known to activate heat shock genes, and found that this did not lead to derepression of \textit{tlpA}. We take this evidence to suggest that the interaction of the promoter with a given TlpA dimer is relatively insensitive to temperature and ignore the temperature sensitivity of $\alpha$.

\begin{figure}[h]
	\includegraphics[width=0.5\textwidth]{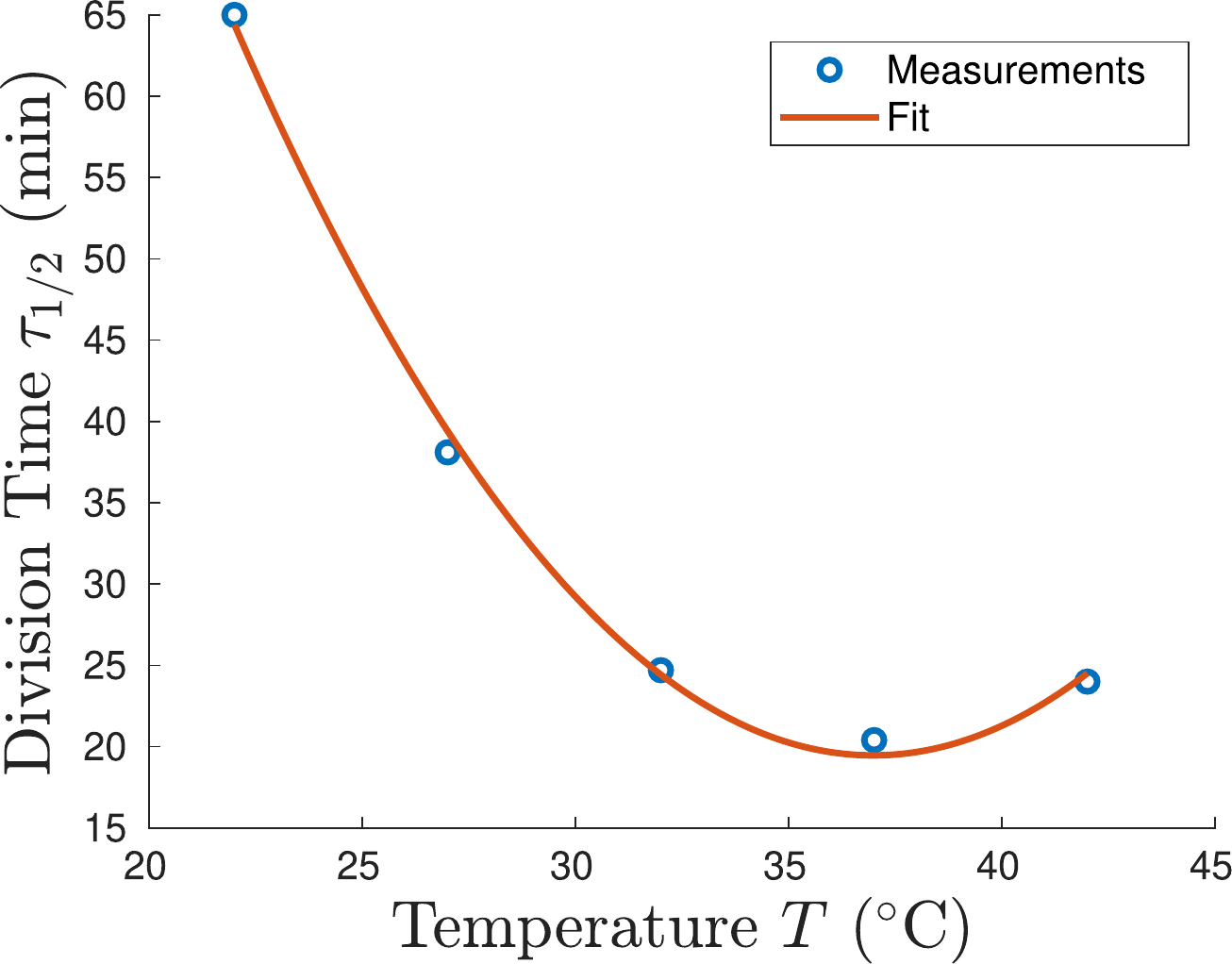}
	\caption{Plot of division time $\tau_{1/2}$ against temperature using data from Ref.\ [37] of the main text. The blue points are the measured average values and the red line is the quadratic fit with least square error.}
	\label{fig:dilutionFit}
\end{figure}

We now consider the temperature dependence of $k^-$, which is set by the cell division time.
Fehlhaber and Kr{\"u}ger measured the temperature dependence of the division time in $\textit{Salmonella enteritidis}$ (Ref.\ [37] of the main text). We fit their data over the range 22-42 ${}^{\circ}$C to a quadratic function using the least squares method, shown in Fig.\ \ref{fig:dilutionFit}.
Using the fit at $T=39{}^{\circ}$C with $k^- = \log(2)/\tau_{1/2}$, we find $(k^-)^{-1}dk^-/dT = - \tau_{1/2}^{-1}d\tau_{1/2}/dT = -0.044 \text{ }({}^{\circ}\text{C})^{-1}$.
To see how this sensitivity compares to that of the fraction $f$ in determining the temperature dependence of the monomer number $\overline{m}$, we recall the expression for $\overline{m}$ in Eq.\ \ref{eq:PDFMeanFull}. Its scaled derivative is
\begin{equation}\label{eq:newDeriv}
		\frac{1}{\overline{m}}\frac{d\overline{m}}{dT} = 
		\left[1-\frac{f(1+\chi)}{2}\right] \frac{1}{f(1-f)} \frac{df}{dT}
		+ \left[\frac{-1-\chi}{2}\right] \frac{1}{k^-} \frac{d k^-}{dT}.
\end{equation}
where the response-like variable $\chi = \sqrt{k^-/[2(1-f)\alpha k^+ +k^-]}$ is positive and less than one.

Now we can compare the magnitudes of the terms containing the temperature sensitivity due to $k^-$ and that due to $f$. In the PD model ($\alpha\rightarrow 0$) at $f=1/2$, the term for $f$ in Eq.\ \ref{eq:newDeriv} is $0.31\text{ }({}^{\circ}\text{C})^{-1}$, while the term for $k^-$ is $0.044\text{ } ({}^{\circ}\text{C})^{-1}$.
In the PDF model,
for the parameters used in the main text ($f=1/2$, $\alpha= 6.94\times 10^{-4}$, $\text{s}^{-1}$, and $k^+/k^-=2521$), the term for $f$ is $0.37\text{ }({}^{\circ}\text{C})^{-1}$, while the term for $k^-$ is $0.035\text{ } ({}^{\circ}\text{C})^{-1}$. In both cases, the term for $f$ is an order of magnitude larger than that for $k^-$. Therefore we neglect the temperature dependence of $k^-$.

\section{Checking the Linear Noise Approximation Using Simulations}
\begin{figure}[h]
	\centering\includegraphics[scale=0.62]{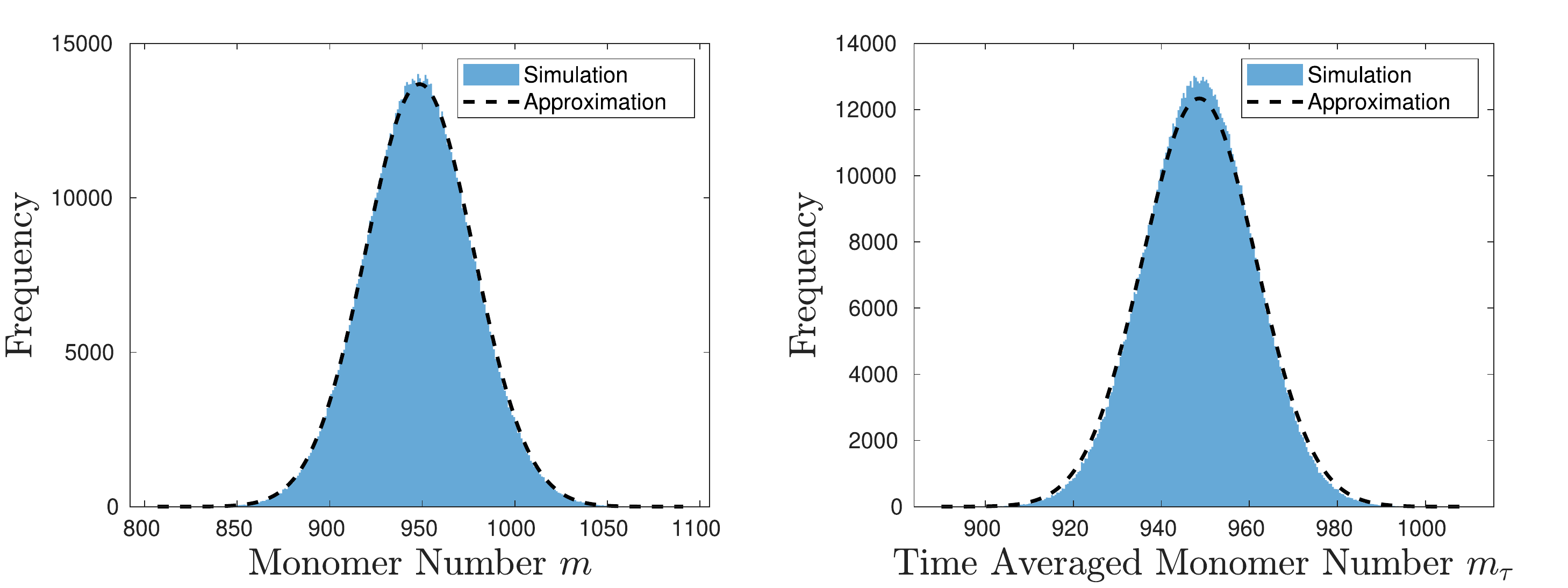}
	\caption{Illustration of the validity of the Gaussian approximation via numerical simulation. The distributions for monomer number (left) and time averaged monomer number (right) over 20 minutes were computed via Gillespie simulations (blue histograms) and the linear noise approximation (black dashed curves).}
\label{fig:gill}
\end{figure}

Employing the Gillespie algorithm (Ref.\ [40] of the main text), we check whether the linear noise approximation assumed above holds for the ``Production-Dilution with Feedback" model. For the simulation, we use the parameters in the main text: $f=1/2$, $\alpha= 6.94\times 10^{-4}$, $k_d=7.1\times 10^{-4}$ $\text{s}^{-1}$, $k^-= 5.5\times 10^{-4}$ $\text{s}^{-1}$, and $k^+/k^-=2521$. Fig.\ \ref{fig:gill} (blue) shows the distribution of monomer numbers (left) and time-averaged monomer numbers over one generation, 20 minutes (right), obtained for $10^6$ simulated trajectories. Both fit well with the linear noise approximation (dashed black) which peaks around the mean $ \overline{m}$ as given in Eq.\ 5 of the main text. We have used the numerical values for this mean $\overline{m}$ and the variance as calculated using Eq.\ \ref{eq:PDFCovar} and Eq.\ \ref{eq:OUTACovar} respectively to generate the plots shown in the left and right panels of Fig.\ \ref{fig:gill}. The theoretically predicted distribution for the time-average has a larger discrepancy with the data than in the case with the instantaneous monomer number. Our theory predicts that the variances for panels A and B are 850.3 and 166.2 respectively, while the data have respective variances of 829.5 and 151.3. We see that the theory predicts a variance that is biased high in both cases. The difference between the predicted variances and the simulated variances are comparable in the two cases, but the discrepancy is relatively larger in the time-averaged case, where the variances are much smaller.

\section{Timescale of Transcriptional Bursts}

When the TlpA dimer is bound to the promoter region for \textit{tlpA}, the gene is not transcribed. If the dimer is unbound, the gene is transcribed and TlpA monomers are produced at rate $k^+$. We introduce $k_{\text{on}}$ and $k_{\text{off}}$, which are the rate constants for a single dimer molecule to bind and unbind from the promoter region respectively. It is straightforward to show that, for a given dimer number $d$, the steady state probability that the gene is off is $(1+k_{\text{on}}d/k_{\text{off}})^{-1}$. For consistency with our deterministic results, we take $k_{\text{on}}/k_{\text{off}}=\alpha$. Therefore, it suffices to estimate either the binding or the unbinding rate. We will estimate the binding rate under the assumption that it is diffusion-limited.
In this case, the binding rate takes the form
\begin{equation}
	k_{\text{on}} = \frac{4\pi D R}{V},
\end{equation}
where $D$ is the relative diffusion coefficient between the promoter and a TlpA dimer, $R$ is the contact radius at which the binding reaction occurs, and $V=1 \text{ }\mu\text{m}^{3}$ is the cell volume (Ref.\ [35] of the main text). 

First, we discuss estimating $R$. We treat the dimer as a sphere whose radius is estimated from its mass and typical values of partial specific volume for proteins (Ref.\ [44] of the main text). The promoter can be enclosed by a sphere whose diameter is the length of the promoter. We assume that the dimer binds to the promoter when it reaches this sphere. This leads to the estimate
\begin{equation}
	R = R_{d} + \frac{L}{2},
\end{equation}
where $R_d$ is the estimated radius of a TlpA dimer and $L$ is the length of the promoter.

Now we turn to estimating the relative diffusion coefficient. Since the promoter region is tethered to the rest of the DNA, we assume that its fluctuations in position are small compared to the excursions of a given dimer molecule. This means that we may approximate the relative diffusion coefficient by the diffusion coefficient of a dimer molecule. We then use the Einstein fluctuation relation to connect the diffusion coefficient to the drag coefficient of the dimer and Stokes' law to connect the drag coefficient to its size. This leads to
\begin{equation}
	D = \frac{k_B T}{6\pi \eta R_d},
\end{equation}
where $\eta$ is the dynamic viscosity of cytosol. 

Putting everything together, we find
\begin{equation}
	k_{\text{on}} = \frac{2}{3} \frac{k_B T}{\eta V} \left[1+ \frac{L}{2 R_d}\right].
\end{equation}
Koski et al.\ (Ref.\ [14] of the main text) state that the mass of a TlpA monomer is 43 kDa, so the dimer has a mass of 86 kDa. This leads to an estimate of $R_d\approx 2.9$ nm (Ref.\ [44] of the main text). Typical transcription factor binding sites are typically $\sim 10$ base pairs long (Ref.\ [45] of the main text), and each base pair is around $0.34$ nm, so this leads to $L\approx 3.4$ nm. The dynamic viscosity of cytosol may be approximated by that of water, which is $6.6\times 10^{-4}$ $\text{kg}/(\text{m}\cdot\text{s})$ at 39 ${}^{\circ}$C. This leads to the estimates $k_{\text{on}} = 6.9 \text{ s}^{-1}$ and $k_{\text{off}} = k_{\text{on}}/\alpha = 9.9\times 10^{3} \text{ s}^{-1}$. Using the parameters in the main text ($f=1/2$, $\alpha= 6.94\times 10^{-4}$, $k_d=7.1\times 10^{-4}$ $\text{s}^{-1}$, $k^-= 5.5\times 10^{-4}$ $\text{s}^{-1}$, and $k^+/k^-=2521$), the typical rate of the receptor switching on is $k_{\text{on}} \overline{d} = 3.3\times 10^{3} \text{ s}^{-1}$. In contrast, the dilution rate is $k^- = 5.5 \times 10^{-4}$ s$^{-1}$, and the dimerization rate is $2 k_d \overline{m} =1.3 \text{ s}^{-1}$. There is a very clear separation of timescales here. The monomer and dimer will effectively respond to the average promoter state. Therefore, we expect the effect of promoter fluctuations on the variance of the monomer and dimer to be negligible, and this is consistent with what we have observed in simulations (data not shown).

\section{Estimate of Thermosensing Precision from Miller Assay Experiments}

We look at the error in temperature sensing inferred from Hurme's measurements of the Miller units at different temperatures (Ref.\ [16] of the main text). In the Miller assay, a promoter is added after the gene of interest. Whenever the target gene is transcribed, the reporter is as well. After some time, optical measurements are taken to quantify gene activity. The result is proportional to the number of times the reporter mRNA has been translated normalized by the cell density (Refs.\ [46] and [47] of the main text).

The data were taken at two different temperatures $T_1$ and $T_2$. We are interested in the activity at intermediate temperatures $T_1<T<T_2$. Let us say that the measured values of the activity $A$ at $T_1$ and $T_2$ are $\overline{A}_{1} \pm\sigma_{A,1}$ and $\overline{A}_{2}\pm\sigma_{A,2}$ respectively. We use linear interpolation to estimate the mean and standard deviation
\begin{gather}
	\overline{A}(T)=\left(\frac{\overline{A}_{2}-\overline{A}_{1}}{T_2-T_1}\right)(T-T_1)
	+\overline{A}_1, \label{eq:MillerMean}\\
	\sigma_A(T)=\left(\frac{\sigma_{A,2}-\sigma_{A,1}}{T_2-T_1}\right)(T-T_1)
	+\sigma_{A,1}.
\end{gather}

We assume that the error is given by linear error propagation. Specifically, the units are measured and Eq.\ \ref{eq:MillerMean} is inverted to solve for temperature $\hat{T} = f(A)$. Linear error propagation at the transition temperature gives the fluctuations in the temperature estimate
\begin{equation}
	\sigma(\hat{T}) = \frac{\sigma_A(T_M)}{|d\overline{A}/dT|} = \sigma_A(T_M) \frac{|T_2-T_1|}{|\overline{A}_{2}-\overline{A}_{1}|},
\end{equation}
which leads to a relative error
\begin{equation}
	\frac{\sigma(\hat{T}) }{|T_2-T_1|}=\frac{\sigma(\hat{T}) }{\Delta T}=\frac{\sigma_A(T_M)}{|\overline{A}_{2}-\overline{A}_{1}|}.
\end{equation}
For \textit{S. typhimurium} molecule $A$ is TlpA, and the experiments were done at $T_1=37^{\circ}C$ and $T_2=43^{\circ}C$ and found $\overline{A}_{1}=68$, $\sigma_{A,1}=16$, $\overline{A}_{2}=294$, $\sigma_{A,2}=116$, $T_M=39^{\circ} C$ (Ref.\ [16] of the main text). This leads to a relative error of 24\%, as stated in the main text.

\section{Including the Miller Assay Reporter in the Theory}
\subsection{Theoretical Calculation}

As mentioned above, the Miller units are proportional to the number of reporter molecules produced per cell. Therefore, we add the production of a reporter molecule $\beta$ to our model. This is produced whenever the monomer would be produced. Since we care about the number produced per cell per generation (around $\tau =20$ minutes), we neglect the effect of degradation or dilution on $\beta$ and start with $\beta_0=0$. We now compute the error in temperature sensing due to the reporter copy number. 

We perform the second-order Kramers-Moyal expansion and then apply the linear noise approximation to the monomer production rate to find
\begin{equation}
d\beta_t = \frac{k^+}{1+\alpha d_t}dt + \sqrt{\frac{k^+}{1+\alpha d_t}} dW^+_t \approx \left[\frac{k^+}{1+\alpha \overline{d}}-\frac{\alpha k^+ \delta d_t}{(1+\alpha \overline{d})^2}\right]dt + \sqrt{\frac{k^+}{1+\alpha \overline{d}}} dW^+_t,
\end{equation}
where $W^+_t$ is a Wiener process with variance $t$. This can be solved by integrating
\begin{equation}\label{eq:betaSol}
	\beta_{\tau} -\frac{k^+ \tau}{1+\alpha \overline{d}} = -\frac{\alpha k^+}{(1+\alpha \overline{d})^2} \int_{0}^{\tau} \delta d_{t'} dt' +\sqrt{\frac{k^+}{1+\alpha \overline{d}}} \int_0^{\tau}dW_{t'}^+.
\end{equation}
Since both of the terms on the right have zero mean, we see that
\begin{equation}
	\overline{\beta}_{\tau} = \frac{k^+ \tau}{1+\alpha \overline{d}}.
\end{equation}
Differentiating with respect to temperature gives
\begin{equation}\label{eq:dbdT1}
	\left| \frac{d\overline{\beta}_{\tau}}{dT} \right| =  \frac{\alpha k^+ \tau}{(1+\alpha \overline{d})^2} \left| \frac{d\overline{d}}{dT} \right|.
\end{equation}
We evaluate this by using Eqs.\ \ref{eq:sigmoidFit}, \ref{eq:DforMF}, \ref{eq:PDFMeanFull}, and \ref{eq:newDeriv}.

Now we need to solve for the fluctuations. Some care is required here, since the noise in $\beta$ is coupled to the noise in the monomer. This follows from decomposing the production-degradation noise in Eq.\ \ref{eq:KMSDEs} as
\begin{equation}
	\sqrt{\frac{k^+}{1+\alpha d_t}+k^- m_t} dW^{(3)}_t = \sqrt{\frac{k^+}{1+\alpha d_t}}dW^+_t - \sqrt{k^- m_t}dW^-_t,
\end{equation}
where $W^+_t$ and $W^-_t$ are independent Wiener processes with variance $t$. Squaring Eq.\ \ref{eq:betaSol} and taking the expectation value gives
\begin{equation}
\sigma^2(\beta_{\tau}) = \frac{(\alpha k^+)^2}{(1+\alpha \overline{d})^4} \int_{0}^{\tau} \mathcal{C}_{1,1}(t_1-t_2)dt_1 dt_2+\frac{k^+}{1+\alpha \overline{d}} \int_0^{\tau}\left\langle dW^+_{t_1} dW^+_{t_2} \right\rangle -\frac{2\alpha (k^+)^{3/2}}{(1+\alpha \overline{d})^{5/2}} \int_0^{\tau} \left\langle dW^+_{t_1} \delta d_{t_2} \right\rangle dt_2.
\end{equation}
By multiplying and dividing the first term by $\tau^2$, it can be written in terms of time averaged covariances from Eq.\ \ref{eq:OUTACovar}. The second term is straightforward to evaluate from the It{\^o} isometry (Ref.\ [30] of the main text). Carrying both of these steps out gives
\begin{equation}\label{eq:betaFirst2}
\sigma^2(\beta_{\tau}) = \frac{(\alpha k^+ \tau)^2}{(1+\alpha \overline{d})^4} [\mathcal{C}_{\text{TA}}(\tau)]_{1,1}+\frac{k^+ \tau}{1+\alpha \overline{d}} -\frac{2\alpha (k^+)^{3/2}}{(1+\alpha \overline{d})^{5/2}} \int_0^{\tau} \left\langle dW^+_{t_1} \delta d_{t_2} \right\rangle dt_2.
\end{equation}

We can evaluate the remaining term using the analytic solution for the Ornstein-Uhlenbeck process from Eq.\ \ref{eq:OUSol}. After linearizing Eq.\ \ref{eq:KMSDEs} around the steady state mean, we have $\vec{\mu}=0$, so there are two terms in $\delta d_t$. The first is the exponential decay of the initial conditions. Since the initial conditions are uncorrelated with the stochastic driving term, this vanishes. The second term, arising from the stochasticity and damping, will make a non-vanishing contribution. It will be convenient to express the noise terms for each reaction as a vector $\vec{W}_t = (W^{(1)}_t, W^{(2)}_t, W^+_t,W^-_t)$. This can be related to $\vec{N}$ in Eq.\ \ref{eq:NNoise} by introducing a matrix
\begin{equation}
	\mathcal{B} = \begin{bmatrix}
		\sqrt{k_d\overline{m}^2 + k_m\overline{d}} & \sqrt{k^- \overline{d}} & 0 & 0 \\
		-2\sqrt{k_d\overline{m}^2 + k_m\overline{d}} & 0 & \sqrt{\dfrac{k^+}{1+\alpha \overline{d}}} & - \sqrt{k^- \overline{m}}
	\end{bmatrix}
\end{equation}
so that $\vec{N}_t=\mathcal{B}\vec{W}_t$. We have
\begin{equation}
 \int_0^{\tau} \left\langle dW^+_{t_1} \delta d_{t_2} \right\rangle dt_2 =  \int_0^{\tau} \left\langle dW^+_{t_1} dt_2 \int_0^{t_2} [e^{\mathcal{J}(t_2-t_3)}\mathcal{B}d\vec{W}_{t_3}]_1\right\rangle.
\end{equation}
Carrying out the matrix multiplication and using the It{\^o} isometry again gives
\begin{equation}
	\int_0^{\tau} \left\langle dW^+_{t_1} \delta d_{t_2} \right\rangle dt_2 = \sqrt{\frac{k^+}{1+\alpha \overline{d}}} \int_0^{\tau} dt' \int_0^{t'} [e^{\mathcal{J}(t'-t'')}]_{1,2}dt'',
\end{equation}
where $\mathcal{J}$ is the Jacobian from Eq.\ \ref{eq:simpJac}. Again, we perform the change of variables $(t',t'')\mapsto (\Delta,t'')$, with $\Delta = t'-t''$. Switching the order of integration, integrating over $t''$ first, and then performing integration by parts for $\Delta$ gives
\begin{equation}
	\int_0^{\tau} \left\langle dW^+_{t_1} \delta d_{t_2} \right\rangle dt_2 = \sqrt{\frac{k^+}{1+\alpha \overline{d}}} \left[-\tau \mathcal{J}^{-1} + \mathcal{J}^{-2}\left[e^{\mathcal{J}\tau}-\mathbb{I}\right]\right]_{1,2}.
\end{equation}
Combining this with the previous two parts from Eq.\ \ref{eq:betaFirst2} to find
\begin{equation}
\sigma^2(\beta_{\tau}) = \frac{(\alpha k^+ \tau)^2}{(1+\alpha \overline{d})^4} [\mathcal{C}_{\text{TA}}(\tau)]_{1,1}+\frac{k^+ \tau}{1+\alpha \overline{d}} -\frac{2\alpha (k^+)^{2}}{(1+\alpha \overline{d})^{3}} \left[-\tau \mathcal{J}^{-1} + \mathcal{J}^{-2}\left[e^{\mathcal{J}\tau}-\mathbb{I}\right]\right]_{1,2}.
\end{equation}
The relative error for temperature sensing in this strategy is $\sigma(\beta_{\tau})/(\Delta T|d\overline{\beta}_{\tau}/dT|)$. For the other parameters, we use the values in the main text: $f=1/2$, $\alpha= 6.94\times 10^{-4}$, $k_d=7.1\times 10^{-4}$ $\text{s}^{-1}$, $k^-= 5.5\times 10^{-4}$ $\text{s}^{-1}$, and $k^+/k^-=2521$. This leads to a relative error of $6.7\%$.

\subsection{Including Translational Bursts}
The transcripts for TlpA and the reporter could have different burst sizes, whose respective means we denote by $\overline{b}_T$ and $\overline{b}_{\beta}$. The terms in the deterministic rate equations take the form (propensity)$\times$(mean change). When adding bursts to TlpA, the mean change in molecule number per reaction changes $1\mapsto \overline{b}_T$. To preserve the mean amount of TlpA, and our consistency with experimental measurements, we map $k^+\mapsto k^+/\overline{b}_T$. Note that this does not generally preserve the amount of $\beta$ in each cell, as the mean production of $\beta$ transforms as
\begin{equation}
	\frac{k^+}{1+\alpha \overline{d}} \mapsto \frac{k^+ \overline{b}_{\beta}/\overline{b}_T}{1+\alpha \overline{d}}.
\end{equation}
Intuitively, the mean burst size of $\beta$ has been measured and is held fixed, but varying $\overline{b}_T$ tunes the frequency of bursts. We take $\overline{b}_{\beta}=7.8$, which is the measured value for the reporter beta-galactosidase (Ref.\ [43] of the main text), and consider multiple values for the mean burst size of TlpA that are typical for bacteria: $\overline{b}_T=1,5,7.8$ and 10.
To find the variance in the amount of $\beta$, we use Gillespie simulations (Ref.\ [40] of the main text) where each production event produces $b_T$ monomers and $b_{\beta}$ reporter molecules, where $b_T$ and $b_{\beta}$ are independent geometric random variables with respective means $\overline{b}_T$ and $\overline{b}_{\beta}$. The derivative is computed using the deterministic result of the previous section, but caution must be exercised when modifying $k^+$, since $\overline{d}$ does not change with the $\overline{b}_T$, but $\overline{\beta}_{\tau}$ does. Modifying Eq.\ \ref{eq:dbdT1}, the result is
\begin{equation}
	\left| \frac{d\overline{\beta}_{\tau}}{dT} \right| =  \frac{\alpha k^+ \tau (\overline{b}_{\beta}/\overline{b}_T)}{(1+\alpha \overline{d})^2} \left| \frac{d\overline{d}}{dT} \right|,
\end{equation}
where $\overline{d}$ is computed according to Eqs.\ \ref{eq:DforMF} and \ref{eq:PDFMeanFull} without changing or mapping $k^+$, as the mapping leaves the production term in $\dot{\overline{m}}$ unchanged. We use the values in the main text: $f=1/2$, $\alpha= 6.94\times 10^{-4}$, $k_d=7.1\times 10^{-4}$ $\text{s}^{-1}$, $k^-= 5.5\times 10^{-4}$ $\text{s}^{-1}$, and $k^+/k^-=2521$. We found that the relative error increased with $\overline{b}_T$, as the mean $\overline{\beta}_{\tau}$ decreased. Over the physiologically relevant range of $5\leq \overline{b}_T\leq 10$, we found that the relative error increased from 23\% to 32\%. Because the range $5\leq \overline{b}_T\leq 10$ is approximate, we round this error range to one significant digit, 20\% to 30\%, as stated in the main text.


\begin{thebibliography}{10}

\bibitem{mccarty_dnak_1991}
J.~S. McCarty and G.~C. Walker.
\newblock {DnaK} as a thermometer: threonine-199 is site of autophosphorylation
  and is critical for {ATPase} activity.
\newblock {\em Proceedings of the National Academy of Sciences},
  88(21):9513--9517, November 1991.

\bibitem{falconi_thermoregulation_1998}
Maurizio Falconi, Bianca Colonna, Gianni Prosseda, Gioacchino Micheli, and
  Claudio~O. Gualerzi.
\newblock Thermoregulation of {Shigella} and {Escherichia} coli {EIEC}
  pathogenicity. {A} temperature-dependent structural transition of {DNA}
  modulates accessibility of {virF} promoter to transcriptional repressor
  {H}-{NS}.
\newblock {\em The EMBO Journal}, 17(23):7033--7043, December 1998.

\bibitem{maeda_effect_1976}
K.~Maeda, Y.~Imae, J.~I. Shioi, and F.~Oosawa.
\newblock Effect of temperature on motility and chemotaxis of {Escherichia}
  coli.
\newblock {\em Journal of Bacteriology}, 127(3):1039--1046, September 1976.

\bibitem{schumann_thermosensors_2007}
Wolfgang Schumann.
\newblock Thermosensors in eubacteria: role and evolution.
\newblock {\em Journal of Biosciences}, 32(3):549--557, April 2007.

\bibitem{klinkert_microbial_2009}
Birgit Klinkert and Franz Narberhaus.
\newblock Microbial thermosensors.
\newblock {\em Cellular and Molecular Life Sciences}, 66(16):2661--2676, August
  2009.

\bibitem{mandin_feeling_2020}
Pierre Mandin and Jörgen Johansson.
\newblock Feeling the heat at the millennium: {Thermosensors} playing with
  fire.
\newblock {\em Molecular Microbiology}, 113(3):588--592, 2020.

\bibitem{berg_physics_1977}
H.C. Berg and E.M. Purcell.
\newblock Physics of chemoreception.
\newblock {\em Biophysical Journal}, 20(2):193--219, November 1977.

\bibitem{endres_accuracy_2008}
Robert~G. Endres and Ned~S. Wingreen.
\newblock Accuracy of direct gradient sensing by single cells.
\newblock {\em Proceedings of the National Academy of Sciences},
  105(41):15749--15754, October 2008.

\bibitem{mora_limits_2010}
Thierry Mora and Ned~S. Wingreen.
\newblock Limits of {Sensing} {Temporal} {Concentration} {Changes} by {Single}
  {Cells}.
\newblock {\em Physical Review Letters}, 104(24):248101, June 2010.

\bibitem{mora_physical_2015}
Thierry Mora.
\newblock Physical {Limit} to {Concentration} {Sensing} {Amid} {Spurious}
  {Ligands}.
\newblock {\em Physical Review Letters}, 115(3):038102, July 2015.

\bibitem{beroz_physical_2017}
Farzan Beroz, Louise~M. Jawerth, Stefan Münster, David~A. Weitz, Chase~P.
  Broedersz, and Ned~S. Wingreen.
\newblock Physical limits to biomechanical sensing in disordered fibre
  networks.
\newblock {\em Nature Communications}, 8(1):16096, July 2017.

\bibitem{fancher_precision_2020}
Sean Fancher, Michael Vennettilli, Nicholas Hilgert, and Andrew Mugler.
\newblock Precision of {Flow} {Sensing} by {Self}-{Communicating} {Cells}.
\newblock {\em Physical Review Letters}, 124(16):168101, April 2020.

\bibitem{dusenbery_limits_1988}
David~B. Dusenbery.
\newblock Limits of thermal sensation.
\newblock {\em Journal of Theoretical Biology}, 131(3):263--271, April 1988.

\bibitem{koski_new_1992}
P.~Koski, H.~Saarilahti, S.~Sukupolvi, S.~Taira, P.~Riikonen, K.~Osterlund,
  R.~Hurme, and M.~Rhen.
\newblock A new alpha-helical coiled coil protein encoded by the {Salmonella}
  typhimurium virulence plasmid.
\newblock {\em Journal of Biological Chemistry}, 267(17):12258--12265, June
  1992.

\bibitem{hurme_dna_1996}
Reini Hurme, Kurt~D. Berndt, Ellen Namork, and Mikael Rhen.
\newblock {DNA} {Binding} {Exerted} by a {Bacterial} {Gene} {Regulator} with an
  {Extensive} {Coiled}-coil {Domain}.
\newblock {\em Journal of Biological Chemistry}, 271(21):12626--12631, May
  1996.

\bibitem{hurme_proteinaceous_1997}
Reini Hurme, Kurt~D Berndt, Staffan~J Normark, and Mikael Rhen.
\newblock A {Proteinaceous} {Gene} {Regulatory} {Thermometer} in {Salmonella}.
\newblock {\em Cell}, 90(1):55--64, July 1997.

\bibitem{fox_gaussian_1978}
Ronald~Forrest Fox.
\newblock Gaussian stochastic processes in physics.
\newblock {\em Physics Reports}, 48(3):179--283, December 1978.

\bibitem{supp}
See Supplemental Material.

\bibitem{servant2000rhea}
Pascale Servant, Cosette Grandvalet, and Philippe Mazodier.
\newblock The rhea repressor is the thermosensor of the hsp18 heat shock
  response in streptomyces albus.
\newblock {\em Proceedings of the National Academy of Sciences},
  97(7):3538--3543, 2000.

\bibitem{jiang_mechanism_2009}
Lili Jiang, Qi~Ouyang, and Yuhai Tu.
\newblock A {Mechanism} for {Precision}-{Sensing} via a {Gradient}-{Sensing}
  {Pathway}: {A} {Model} of {Escherichia} coli {Thermotaxis}.
\newblock {\em Biophysical Journal}, 97(1):74--82, July 2009.

\bibitem{paulick_thermo_2017}
Anja Paulick, Vladimir Jakovljevic, SiMing Zhang, Michael Erickstad, Alex
  Groisman, Yigal Meir, William~S Ryu, Ned~S Wingreen, and Victor Sourjik.
\newblock Mechanism of bidirectional thermotaxis in \textit{Escherichia coli}.
\newblock {\em eLife}, 6:e26607, August 2017.

\bibitem{note_feedback}
In the high temperature response, the negative feedback is due to
  transcriptional repression, for example via dimers repressing monomer
  production as discussed herein; in the heat shock response, it is due to
  proteases degrading or chaperones conformationally changing the oligomers
  that form \cite{mccarty_dnak_1991}; in {\it E.\ coli} thermotaxis, it is due
  to the methylation of the receptors
  \cite{jiang_mechanism_2009,paulick_thermo_2017}.

\bibitem{note_absolute}
In response to a temperature increase, TlpA levels remain high for at least
  two hours \cite{hurme_proteinaceous_1997}, and TlpA production remains
  constant for at least 24 hours \cite{piraner_tunable_2017}, whereas other
  heat shock factors such as $\sigma^{32}$ decrease in level 20 minutes after
  induction \cite{zhao2005global}.

\bibitem{piraner_tunable_2017}
Dan~I. Piraner, Mohamad~H. Abedi, Brittany~A. Moser, Audrey Lee-Gosselin, and
  Mikhail~G. Shapiro.
\newblock Tunable thermal bioswitches for in vivo control of microbial
  therapeutics.
\newblock {\em Nature Chemical Biology}, 13(1):75--80, January 2017.

\bibitem{piraner_modular_2019}
Dan~I. Piraner, Yan Wu, and Mikhail~G. Shapiro.
\newblock Modular {Thermal} {Control} of {Protein} {Dimerization}.
\newblock {\em ACS Synthetic Biology}, 8(10):2256--2262, October 2019.

\bibitem{heyn_circular_1975}
Maarten~P. Heyn and Wolfgang~O. Weischet.
\newblock Circular dichroism and fluorescence studies on the binding of ligands
  to the α subunit of tryptophan synthase.
\newblock {\em Biochemistry}, 14(13):2962--2968, July 1975.

\bibitem{cover_elements_2012}
Thomas~M Cover and Joy~A Thomas.
\newblock {\em Elements of information theory}.
\newblock John Wiley \& Sons, 2012.

\bibitem{van_kampen_stochastic_2011}
N.~G. Van~Kampen.
\newblock {\em Stochastic {Processes} in {Physics} and {Chemistry}}.
\newblock Elsevier, August 2011.

\bibitem{gardiner_stochastic_2009}
Crispin Gardiner.
\newblock {\em Stochastic {Methods}: {A} {Handbook} for the {Natural} and
  {Social} {Sciences}}.
\newblock Springer Berlin Heidelberg, January 2009.

\bibitem{klebaner_introduction_2012}
Fima~C. Klebaner.
\newblock {\em Introduction {To} {Stochastic} {Calculus} {With}
  {Applications}}.
\newblock World Scientific Publishing Company, March 2012.

\bibitem{note_DeltaT}
A cell must be able to determine temperature changes to a better precision than
  the width of its temperature sensitive region $\Delta T$. For this reason, we
  define relative error with respect to $\Delta T$, not the mean temperature
  $\overline{T}$, in Eq.\ \ref{eq:fixedPool} and thereafter. This is in
  contrast to the case of the perfect instrument (Eq.\ \ref{eq:extrinsic}), for
  which there is no cell-defined temperature-sensitive region. It is worth
  noting that even if $\overline{T}$ were replaced with $\Delta T$ in Eq.\
  \ref{eq:extrinsic}, the relative error would increase by roughly two orders
  of magnitude, which is still far less than that of the biochemical models
  considered in this work.

\bibitem{roob_cooperative_2016}
Edward Roob, Nicola Trendel, Pieter Rein ten Wolde, and Andrew Mugler.
\newblock Cooperative {Clustering} {Digitizes} {Biochemical} {Signaling} and
  {Enhances} its {Fidelity}.
\newblock {\em Biophysical Journal}, 110(7):1661--1669, April 2016.

\bibitem{note_dimer}
The TlpA dimer number was experimentally estimated to be $\overline{d} =
  684$ at $T = 37\ ^{\circ}$C \cite{hurme_proteinaceous_1997}, where $f = 0.2$.
  Because $f = \overline{m}/(\overline{m}+2\overline{d})$, we have
  $\overline{m} = 2\overline{d}f/(1-f) = 342$. In the fixed pool model, this
  implies $n = \overline{m}/f = 1710$. In the production-dilution model, this
  implies $k^+/k^- = \overline{m}/f = 1710$. In the production-dilution with
  feedback model, we insert $\overline{m} = 342$, $f=0.2$, and $\alpha =
  \alpha^* = 1.75k^-/k^+$ into Eq.\ \ref{eq:fullMean} to obtain $k^+/k^- =
  2521$.

\bibitem{chao_use_1998}
Heman Chao, Daisy~L Bautista, Jennifer Litowski, Randall~T Irvin, and Robert~S
  Hodges.
\newblock Use of a heterodimeric coiled-coil system for biosensor application
  and affinity purification.
\newblock {\em Journal of Chromatography B: Biomedical Sciences and
  Applications}, 715(1):307--329, September 1998.

\bibitem{smit_outer_1975}
J.~Smit, Y.~Kamio, and H.~Nikaido.
\newblock Outer membrane of {Salmonella} typhimurium: chemical analysis and
  freeze-fracture studies with lipopolysaccharide mutants.
\newblock {\em Journal of Bacteriology}, 124(2):942--958, November 1975.

\bibitem{code}
The expressions for $\sigma(\hat{T})/\Delta T$ vs.\ $\tau$ for all models
  are calculated analytically by matrix inversion in Mathematica. Code is
  available at \url{https://github.com/BumblingBulblax/ProteinThermometry}.

\bibitem{https://doi.org/10.1046/j.1365-2672.1998.00410.x}
Fehlhaber and Krüger.
\newblock The study of salmonella enteritidis growth kinetics using rapid
  automated bacterial impedance technique.
\newblock {\em Journal of Applied Microbiology}, 84(6):945--949, 1998.

\bibitem{milo_cell_2015}
Ron Milo and Rob Phillips.
\newblock {\em Cell {Biology} by the {Numbers}}.
\newblock Garland Science, December 2015.

\bibitem{lowrie_division_1979}
D.~B. Lowrie, V.~R. Aber, and M.~E. Carrol.
\newblock Division and death rates of {Salmonella} typhimurium inside
  macrophages: use of penicillin as a probe.
\newblock {\em Journal of General Microbiology}, 110(2):409--419, February
  1979.

\bibitem{gillespie_exact_1977}
Daniel~T. Gillespie.
\newblock Exact stochastic simulation of coupled chemical reactions.
\newblock {\em The Journal of Physical Chemistry}, 81(25):2340--2361, December
  1977.

\bibitem{note_alpha}
We find that the relative fluctuations $\sigma(m)/\overline{m}$ scale as
  $\alpha^{1/4}$ and as $c + s \alpha$ for large and small $\alpha$
  respectively, where $c$ and $s$ are independent of $\alpha$. For $0<f<0.77$,
  the slope $s$ is positive. Because we are generally concerned with values of
  $f$ near $1/2$, we conclude that $\sigma(m)/\overline{m}$ increases
  monotonically for all $\alpha$.

\bibitem{raj2008nature}
Arjun Raj and Alexander Van~Oudenaarden.
\newblock Nature, nurture, or chance: stochastic gene expression and its
  consequences.
\newblock {\em Cell}, 135(2):216--226, 2008.

\bibitem{xie2008single}
X~Sunney Xie, Paul~J Choi, Gene-Wei Li, Nam~Ki Lee, and Giuseppe Lia.
\newblock Single-molecule approach to molecular biology in living bacterial
  cells.
\newblock {\em Annu. Rev. Biophys.}, 37:417--444, 2008.

\bibitem{erickson_size_2009}
Harold~P. Erickson.
\newblock Size and {Shape} of {Protein} {Molecules} at the {Nanometer} {Level}
  {Determined} by {Sedimentation}, {Gel} {Filtration}, and {Electron}
  {Microscopy}.
\newblock {\em Biological Procedures Online}, 11(1):32--51, December 2009.

\bibitem{10.1534/genetics.112.143370}
Alexander~J Stewart, Sridhar Hannenhalli, and Joshua~B Plotkin.
\newblock {Why Transcription Factor Binding Sites Are Ten Nucleotides Long}.
\newblock {\em Genetics}, 192(3):973--985, 11 2012.

\bibitem{miller_experiments_1972}
Jeffrey~H. Miller and Jeffrey~B. Miller.
\newblock {\em Experiments in {Molecular} {Genetics}}.
\newblock Cold Spring Harbor Laboratory, 1972.

\bibitem{garcia_comparison_2011}
Hernan~G. Garcia, Heun~Jin Lee, James~Q. Boedicker, and Rob Phillips.
\newblock Comparison and {Calibration} of {Different} {Reporters} for
  {Quantitative} {Analysis} of {Gene} {Expression}.
\newblock {\em Biophysical Journal}, 101(3):535--544, August 2011.

\bibitem{foreman_variability_2020}
Robert Foreman and Roy Wollman.
\newblock Mammalian gene expression variability is explained by underlying cell
  state.
\newblock {\em Molecular Systems Biology}, 16(2):e9146, 2020.

\bibitem{siggia_decisions_2013}
Eric~D. Siggia and Massimo Vergassola.
\newblock Decisions on the fly in cellular sensory systems.
\newblock {\em Proceedings of the National Academy of Sciences},
  110(39):E3704--E3712, September 2013.

\bibitem{berger_statistical_1985}
James~O. Berger.
\newblock {\em Statistical {Decision} {Theory} and {Bayesian} {Analysis}}.
\newblock Springer Science \& Business Media, August 1985.

\bibitem{peskir_optimal_2006}
Goran Peskir and Albert Shiryaev.
\newblock {\em Optimal {Stopping} and {Free}-{Boundary} {Problems}}.
\newblock Springer-Verlag NY, August 2006.

\bibitem{zhao2005global}
Kai Zhao, Mingzhu Liu, and Richard~R Burgess.
\newblock The global transcriptional response of escherichia coli to induced
  $\sigma$32 protein involves $\sigma$32 regulon activation followed by
  inactivation and degradation of $\sigma$32 in vivo.
\newblock {\em Journal of Biological Chemistry}, 280(18):17758--17768, 2005.

\end{thebibliography}
\end{document}